\documentclass[]{bytedance_seed}
\usepackage{amsmath,amssymb,array,longtable,needspace,xurl,placeins}
\usepackage{tabularx,siunitx,caption,multirow}
\DeclareCaptionFont{appendix}{\fontsize{9}{11}\selectfont}
\DeclareCaptionLabelSeparator{appendixcustom}{}
\DeclareCaptionFormat{appendixcustom}{{\fontsize{8.8}{11}\selectfont\sffamily\bfseries #1 #2} #3}
\newcommand{\AppendixTableStyle}{%
  \fontsize{9}{11}\selectfont
  \renewcommand{\arraystretch}{1.14}%
  \setlength{\tabcolsep}{5pt}%
}
\newcommand{\AppendixPresentation}{%
  \captionsetup{font=appendix,format=appendixcustom,labelsep=appendixcustom,
    justification=raggedright,singlelinecheck=false,
    skip=6pt,width=\linewidth}%
  \setlength{\floatsep}{12pt plus 2pt minus 2pt}%
  \setlength{\textfloatsep}{12pt plus 2pt minus 2pt}%
  \setlength{\intextsep}{10pt plus 2pt minus 2pt}%
  \renewcommand{\topfraction}{0.95}%
  \renewcommand{\bottomfraction}{0.90}%
  \renewcommand{\textfraction}{0.05}%
  \renewcommand{\floatpagefraction}{0.80}%
  \setcounter{topnumber}{4}\setcounter{bottomnumber}{3}\setcounter{totalnumber}{6}%
  \setlength{\LTcapwidth}{\linewidth}%
  \setlength{\LTpre}{10pt}\setlength{\LTpost}{10pt}%
  \sisetup{group-digits=none,exponent-product=\times}%
}
\makeatletter
\AtBeginDocument{\setlength{\@fptop}{0pt}}
\makeatother

\DeclareFontShape{T1}{bytesans}{b}{n}{<-> s * [1] seed/bytesans}{}

\setcitestyle{super,sort&compress}

\title{FACTO: fast atom-centered tensor operator for accurate and efficient foundational machine-learned interatomic potentials}

\author[\dagger]{Siyuan Liu}
\author[]{Chang Han}
\author[]{Sheng Gong}
\author[]{Yanping Yang}
\author[]{Tianze Zheng}
\author[]{Yu Xia}
\author[]{Kuang Yu}
\author[]{Shenzhen Xu}
\author[\dagger]{Wen Yan}

\affiliation[]{ByteDance Seed}
\contribution[\dagger]{Corresponding authors}

\abstract{
High-throughput inference remains a central challenge for equivariant machine-learned interatomic potentials.
Here we introduce FACTO, an atom-centered equivariant architecture that confines evolving learned state and dense channel mixing to atoms, expressing neighbor interactions as fully fusible reductions.
We complement this design with custom register-accumulated aggregation kernels and tensor-core FP32 emulation, accelerating inference while retaining FP32-level energy conservation under emulated arithmetic in molecular dynamics.
Comprehensive evaluations on Matbench Discovery, MLIP Arena and MatCalc demonstrate high throughput alongside strong predictive accuracy and physical reliability.
On Matbench Discovery, the four FACTO-OAM variants show top-tier performance by overall rank sum among 46 models with complete scores, advancing the accuracy--throughput Pareto frontier.
On MLIP Arena, FACTO achieves state-of-the-art stability under extreme heating and compression, alongside high fidelity in hydrogen combustion.
On MatCalc, FACTO outperforms the compared baselines in structural and elastic predictions while retaining competitive accuracy for energetic and thermal properties.
}

\date{October 30, 2026}
\correspondence{Siyuan Liu at \email{siyuan.liu@bytedance.com} and Wen Yan at \email{wen.yan@bytedance.com}.}

\begin{document}
\maketitle

\section{Introduction}
\label{sec:introduction}

\begingroup
\clubpenalties 1 10000
\widowpenalties 1 10000

The scientific reach of atomistic simulation depends on the fidelity of an
interatomic potential and the cost of exploring its energy landscape.
Structural relaxation and molecular dynamics require repeated force evaluations,
whose cost limits system size, trajectory length and statistical sampling
under a fixed computational budget.
Machine-learned interatomic potentials (MLIPs) reduce this cost by learning from
density-functional theory (DFT) calculations, extending materials modeling
across chemical compositions and beyond the scales accessible to direct
first-principles simulation\citep{deng2023chgnet}.
When an MLIP drives relaxation or dynamics, each predicted force determines the
subsequent configuration, so force errors can deflect trajectories even when
static test-set errors appear small\citep{fu2023forces}.
Practical atomistic simulation therefore requires joint optimization of
evaluation speed and physical fidelity across the potential-energy surface.

\begin{figure}[!t]
\centering
\includegraphics[width=\linewidth]{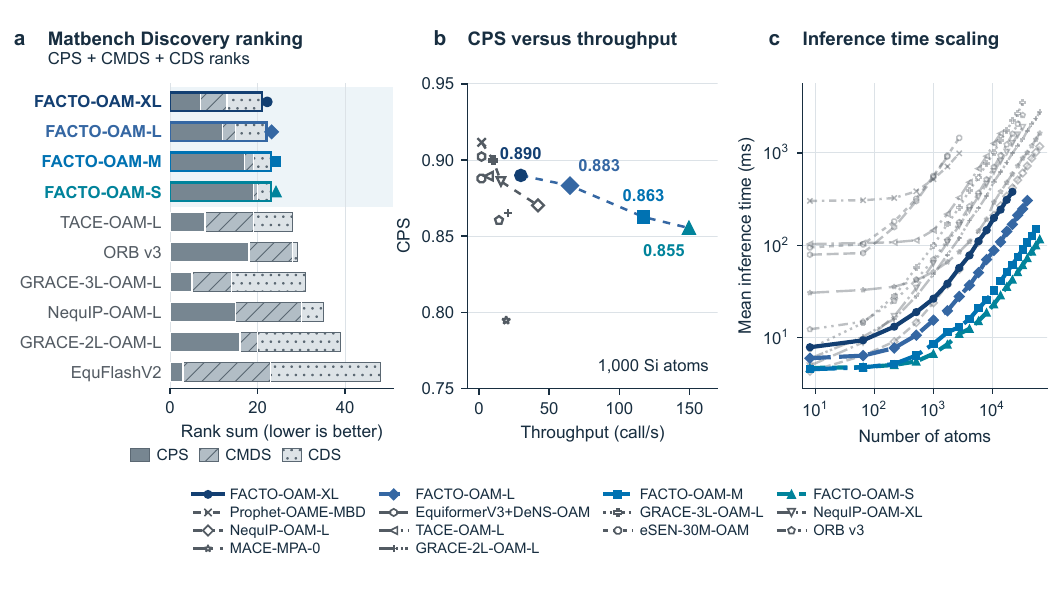}
\caption{{\bfseries Matbench Discovery performance and silicon inference scaling.}
{\bfseries a,} Rank sums across the Matbench Discovery combined
performance (CPS), molecular-dynamics (CMDS) and diatomic (CDS) scores (lower is better).
The ten lowest rank sums among the four FACTO-OAM variants and 42 public models
with complete scores\cite{matbench2026snapshot} are shown (end markers match the shared legend).
{\bfseries b,} CPS versus inference throughput for 1,000-atom silicon supercells.
Numbers give the FACTO-OAM CPS values, and dashed lines trace separate Pareto frontiers
for the FACTO series (blue) and leading public models (grey).
{\bfseries c,} Inference latency versus atom count for silicon supercells on logarithmic axes,
comparing the four FACTO-OAM variants with ten public models. For the FACTO series,
curves trace optimized inference scaling, transitioning from standard FP32 AOTI compilation
at small system sizes to the fully accelerated pipeline beyond the crossover threshold
(see Fig.~\ref{fig:facto-acceleration} and Methods for details).}
\label{fig:facto-overview}
\end{figure}

An MLIP's representation of local environments around atoms shapes both the interactions it
can learn and the operations required to evaluate them.
Equivariant neural networks encode directional information in vector and tensor
features that are exchanged through pairwise edge operations in each
message-passing layer during force evaluation\citep{batzner2022nequip,batatia2022mace,musaelian2023allegro}.
Previous studies have simplified these tensor interactions and moved
equivariant operations onto atoms
\citep{simeon2023tensornet,zhang2026node,reschutzegger2026geodite}, while
complementary work on kernel fusion has reduced the storage of edge
intermediates and associated data movement\citep{lee2025flashtp}.
Yet the model architecture itself limits how much edge memory traffic can be
removed: computationally intensive edge operations inevitably materialize large
temporary edge tensors within each layer, while learned edge embeddings require
persistent state to be carried across layers.
Avoiding both bottlenecks motivates retaining geometric state in atom-wise
features while making all edge operations fusible.

Here we introduce FACTO, the Fast Atom-Centered Tensor Operator, an equivariant
neural network designed to reduce the computational cost and memory traffic of
repeated force evaluations.
Rather than evaluating explicit Clebsch--Gordan tensor products or dense
projections on edges, FACTO encodes directional geometry in atom-centered tensor
features and couples scalar and tensor streams through atom-wise projections
and dot-product contractions.
All evolving learned state resides on atoms, unlike the scalar edge embeddings
propagated between layers in Geodite\citep{reschutzegger2026geodite}.
Once the shared geometric edge encodings are computed, each interaction block
avoids both temporary edge intermediates and cross-layer edge state, leaving
only lightweight, fully fusible pairwise reductions.

Beyond kernel fusion, we reduce message-aggregation overhead with custom
atom-centered kernels that accumulate neighbor contributions without GPU atomic
operations.
Alongside memory-bound neighbor aggregation, inference throughput is governed
by the dense matrix multiplications in atom-wise projections.
Standard lower-precision tensor-core formats increase throughput, but their
rounding errors can degrade energy and force fidelity during dynamics.
We therefore adopt tensor-core FP32 emulation via a three-product bfloat16
decomposition (BF16$\times$3)\citep{henry2019bfloat16} to exploit bfloat16
tensor-core throughput while recovering the leading FP32 residual bits.
Together, these optimizations target the two primary bottlenecks of equivariant
force evaluation: memory-bound neighbor aggregation and compute-bound atom-wise
projections.
Because both patterns recur across modern MLIPs, the optimizations transfer
beyond FACTO.

We evaluate FACTO on Matbench Discovery, MatCalc and MLIP Arena
\citep{riebesell2025matbench,kaplan2025matpes,chiang2025arena} and benchmark
large-system inference scaling and microcanonical dynamics in silicon
to assess physical fidelity alongside computational cost.
The FACTO-OAM family occupies the top four overall rank-sum positions on
Matbench Discovery (Fig.~\ref{fig:facto-overview}a).
Across model sizes, it advances the accuracy--throughput frontier
(Fig.~\ref{fig:facto-overview}b) and maintains efficient inference as silicon
system size grows (Fig.~\ref{fig:facto-overview}c).
On MatCalc, the family outperforms the compared baselines on structural
and elastic benchmarks, with FACTO-OAM-XL also leading formation-energy
accuracy while nearly eliminating off-equilibrium force softening.
On MLIP Arena, it combines tight force--energy conservation with
state-of-the-art survival under extreme heating, high-pressure compression and
reactive hydrogen combustion.
Our silicon benchmarks further confirm that pairing custom neighbor aggregation
with BF16$\times$3 emulation yields compounding speedups for large systems, while
BF16$\times$3 emulation retains strict-FP32-level energy fluctuations in
microcanonical dynamics.
Our main contributions are:
\begin{itemize}
  \item \textit{Atom-centered equivariant architecture with fully fusible pairwise operations.}
  FACTO confines evolving learned state and dense channel mixing to atoms and
  expresses directional couplings through dot-product feedback and factorized
  transport, allowing all within-block pairwise operations to be fused into
  single-pass neighbor reductions.
  \item \textit{Advancing the accuracy--throughput frontier across atomistic benchmarks.}
  Across Matbench Discovery, MatCalc and MLIP Arena, FACTO delivers
  high accuracy spanning crystal discovery, crystalline structural and
  mechanical response, off-equilibrium force fidelity and extreme dynamical
  stability at substantially higher throughput than existing equivariant
  potentials.
  \item \textit{Custom atom-centered aggregation and transferable tensor-core FP32 emulation.}
  We pair register-accumulated neighbor kernels that avoid GPU atomic additions
  in forward neighbor reductions with three-product bfloat16 FP32 emulation (BF16$\times$3) to accelerate
  memory-bound neighbor reductions and compute-bound atom-wise projections,
  with BF16$\times$3 preserving trajectory energy conservation in microcanonical
  dynamics and both optimizations transferring to future MLIPs.
\end{itemize}

\endgroup

\section{Results}

\subsection{The FACTO architecture}
\label{sec:facto-architecture}

\begin{figure}[tp]
  \centering
  \includegraphics[width=\linewidth]{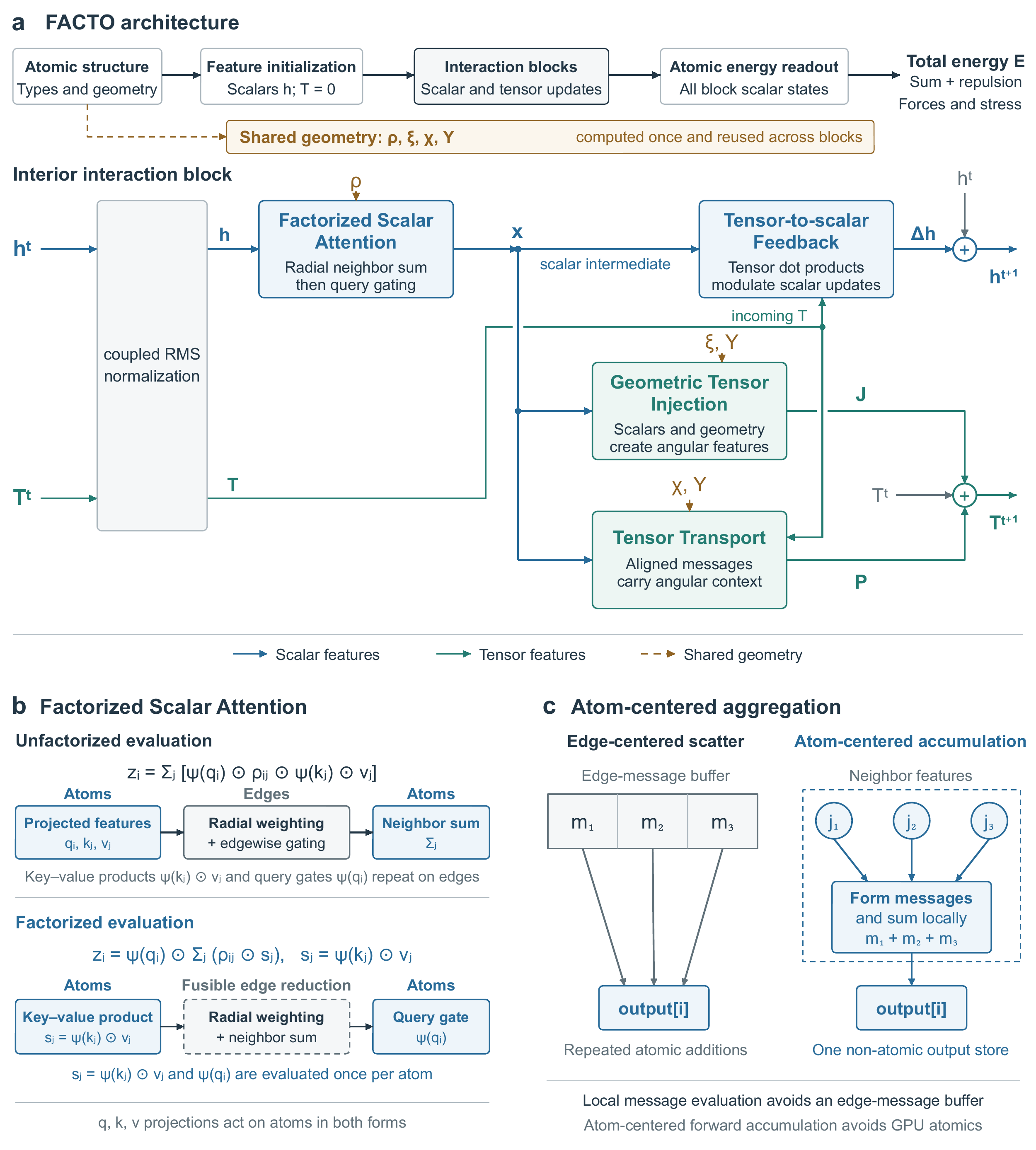}
  \caption{{\bfseries FACTO architecture and atom-centered execution.}
  {\bfseries a,} Coupled scalar and tensor updates in block $t$ with scaled residual
  additions. Shared scalar filters $\boldsymbol\rho$,
  tensor filters $\boldsymbol\xi$, cutoff weights $\chi$ and spherical
  harmonics $Y$ are supplied along dashed arrows.
  {\bfseries b,} Equivalent evaluations of one scalar-attention group using the
  same atom-wise projections $\mathbf q,\mathbf k,\mathbf v$ (group index $g$
  omitted for clarity). Here $\psi$ is SiLU and $\odot$ denotes channelwise
  multiplication. Factorization reduces the edge stage to the dashed, fusible
  neighbor sum.
  {\bfseries c,} Forward aggregation for one output component.
  In the materialized reference, edge messages are stored and accumulated with
  GPU atomic additions (indivisible memory updates);
  in receiver-owned accumulation, they are formed and summed locally before a single
  ordinary output store, without a GPU atomic read--modify--write update.
  Neighbor indices and shared geometry remain in memory; the three neighbors are illustrative.}
  \label{fig:facto-architecture}
\end{figure}

We designed FACTO to keep evolving learned state and dense channel mixing on
atoms, expressing neighbor interactions as fusible reductions
(Fig.~\ref{fig:facto-architecture}a). Within each interaction block, four
operations couple the scalar and tensor streams: Factorized Scalar Attention
aggregates scalar neighbor information to drive both streams, Tensor-to-scalar
Feedback returns accumulated angular information to the scalar update, and
Geometric Tensor Injection and Tensor Transport create and propagate
directional tensor features. Bias terms are omitted from the equations below
for clarity.

\noindent{\bfseries Feature initialization.}
We represent each structure as a periodic neighbor graph in which each atom $i$
of atomic number $Z_i$ carries scalar features
$\mathbf h_i\in\mathbb R^D$ and tensor features
$\mathbf T_i^\ell\in\mathbb R^{(2\ell+1)\times C_\ell}$ across angular orders
$1\leq\ell\leq L$. The channel counts $C_\ell$ can vary with $\ell$; in
particular, we use fewer channels at higher angular orders.
We initialize the scalar state by applying a
learned affine projection and the SiLU activation $\psi$ to fixed elemental
descriptors, whereas the tensor state starts at zero and is populated by the
first application of Geometric Tensor Injection.

Each directed edge from neighbor $j$ to atom $i$ (including periodic images) is
characterized by the displacement vector $\mathbf r_{ij}$, distance
$r_{ij}=\|\mathbf r_{ij}\|$ and unit direction
$\widehat{\mathbf r}_{ij}=\mathbf r_{ij}/r_{ij}$. We expand the interatomic
distance relative to the interaction radius $R_{\rm c}$ using fixed Gaussian
smearing:
\begin{equation}
  u_{ij,n}
    =\exp\!\left[-\frac{(r_{ij}/R_{\rm c}-\mu_n)^2}{2s_r^2}\right],
  \label{eq:facto-radial-basis}
\end{equation}
where the centers $\mu_n$ and width $s_r$ are fixed.
To smoothly limit the interaction range and suppress learned contributions at
very short distances, we combine $C^3$-smooth polynomial cutoffs $f_R$ at
$R_{\rm c}$ and a short-range radius $R_{\rm s}$:
\begin{equation}
  \begin{aligned}
    f_{R}(r)
      &=\begin{cases}
        1-84\left(\frac{r}{R}\right)^6
        +216\left(\frac{r}{R}\right)^7
        -189\left(\frac{r}{R}\right)^8
        +56\left(\frac{r}{R}\right)^9,
          &0\leq r<R,\\
        0,&r\geq R,
      \end{cases}\\
    \chi_{ij}
      &=f_{R_{\rm c}}(r_{ij})[1-f_{R_{\rm s}}(r_{ij})].
  \end{aligned}
  \label{eq:facto-cutoff-envelope}
\end{equation}
We multiply the Gaussian basis $\mathbf u_{ij}$ by the cutoff envelope to obtain
the shared radial features $\boldsymbol\phi_{ij}=\chi_{ij}\mathbf u_{ij}$.
We project these features directly, without a radial multilayer perceptron or
pair-type embeddings, to obtain the scalar filter $\boldsymbol\rho_{ij}$ for
Factorized Scalar Attention and the tensor filters $\boldsymbol\xi_{ij}^\ell$
for Geometric Tensor Injection:
\begin{equation}
  \boldsymbol\phi_{ij}
    =\chi_{ij}\mathbf u_{ij},
  \qquad
  \boldsymbol\rho_{ij}
    =\chi_{ij}\psi(W_\rho\boldsymbol\phi_{ij}),
  \qquad
  \boldsymbol\xi_{ij}^\ell
    =W_\xi^\ell\boldsymbol\phi_{ij},
  \label{eq:facto-radial-filters}
\end{equation}
where the outer envelope factor on $\boldsymbol\rho_{ij}$ ensures an exact zero
cutoff when the linear projection includes a bias term (whereas $W_\xi^\ell$ is
bias-free).
Separately from the radial filters, we evaluate real spherical harmonics
$Y_{\ell m}(\widehat{\mathbf r}_{ij})$ from the unit direction
$\widehat{\mathbf r}_{ij}$ to encode edge orientation.
Both the radial filters and spherical harmonics are
computed once per forward pass and reused across all interaction blocks,
keeping radial projections outside the repeated message-passing loop.

\noindent{\bfseries Coupled RMS normalization.}
At the start of block $t$, we normalize the stored states
$\mathbf h_i^{(t)}$ and $\mathbf T_i^{(t,\ell)}$ to obtain the block inputs
$\mathbf h_i$ and $\mathbf T_i^\ell$ used in the operations below:
\begin{equation}
  \begin{aligned}
    \nu_i
      &=\frac{1}{D}\sum_{a=1}^D(h_{i,a}^{(t)})^2,
    &\omega_{i,c}^\ell
      &=\frac{1}{2\ell+1}\sum_{m=-\ell}^{\ell}(T_{i,mc}^{(t,\ell)})^2,\\
    h_{i,a}
      &=\frac{\gamma_a^h h_{i,a}^{(t)}}{\sqrt{\nu_i+\epsilon}},
    &T_{i,mc}^\ell
      &=\frac{\gamma_c^\ell T_{i,mc}^{(t,\ell)}}
        {\sqrt{\nu_i+\omega_{i,c}^\ell+\epsilon}},
  \end{aligned}
  \label{eq:facto-coupled-rms}
\end{equation}
where $a$ and $c$ index channels, $m$ indexes angular components, $\gamma_a^h$
and $\gamma_c^\ell$ are learned per-channel gains, and $\epsilon>0$ stabilizes
the denominators. Including the scalar mean square $\nu_i$ in the tensor
denominator provides a finite normalization scale even when angular features
vanish in highly symmetric local environments.

\noindent{\bfseries Factorized Scalar Attention.}
To aggregate scalar neighbor information across $G$ channel groups of width
$D/G$, we first apply atom-wise linear projections to obtain per-group queries
$\mathbf q_i^g=W_q^g\mathbf h_i$ on atom $i$ and shared keys and values
$\mathbf k_j=W_k\mathbf h_j$ and $\mathbf v_j=W_v\mathbf h_j$ on neighbor $j$.
In each group $g$, we couple the query gate $\psi(\mathbf q_i^g)$, the
radial filter $\boldsymbol\rho_{ij}^g$, and the key--value product
$\psi(\mathbf k_j)\odot\mathbf v_j$. Because the query gate depends only on
atom $i$ and the key--value product depends only on atom $j$, we precompute
$\mathbf s_j=\psi(\mathbf k_j)\odot\mathbf v_j$ on atoms and factor
$\psi(\mathbf q_i^g)$ outside the neighbor sum
(Fig.~\ref{fig:facto-architecture}b):
\begin{equation}
  \begin{aligned}
    \mathbf z_i^g
      &=\sum_{j\in\mathcal N(i)}
        \psi(\mathbf q_i^g)\odot\boldsymbol\rho_{ij}^g
        \odot\psi(\mathbf k_j)\odot\mathbf v_j\\
      &=\psi(\mathbf q_i^g)\odot
        \sum_{j\in\mathcal N(i)}\boldsymbol\rho_{ij}^g\odot\mathbf s_j,
        \qquad\mathbf s_j=\psi(\mathbf k_j)\odot\mathbf v_j,\\
    \mathbf x_i
      &=\psi\!\left(W_x[\mathbf z_i^1;\cdots;\mathbf z_i^G]\right),
  \end{aligned}
  \label{eq:facto-scalar-attention}
\end{equation}
where $\mathcal N(i)$ is the neighbor set of atom $i$, $\odot$ denotes
channelwise multiplication and brackets denote channel concatenation. Rather
than repeating $\psi(\mathbf k_j)\odot\mathbf v_j$ and $\psi(\mathbf q_i^g)$ on
every edge, we evaluate the key--value product $\mathbf s_j$ and query gate
$\psi(\mathbf q_i^g)$ once per atom and apply $\psi(\mathbf q_i^g)$ after
aggregation, leaving only radial weighting and summation on the edges.

\noindent{\bfseries Tensor-to-scalar Feedback.}
Tensor-to-scalar Feedback incorporates accumulated angular information into the
scalar stream. For each angular order $\ell$, we apply atom-wise channel
projections $U_\ell$ and $V_\ell$ to the normalized tensor state and contract
over the angular index $m$ via componentwise dot products:
\begin{equation}
  \begin{aligned}
    \mathbf d_i^\ell
      &=\sum_{m=-\ell}^{\ell}
        (\mathbf T_i^\ell U_\ell)_m
        \odot(\mathbf T_i^\ell V_\ell)_m,\\
    \mathbf f_i&=W_f[\mathbf d_i^1;\cdots;\mathbf d_i^L],\\
    \Delta\mathbf h_i
      &=\mathbf f_i\odot(W_1\mathbf x_i)+W_2\mathbf x_i,
  \end{aligned}
  \label{eq:facto-tensor-feedback}
\end{equation}
where a subscript $m$ selects the channel vector corresponding to angular
component $m$ (with channel indices suppressed). Because these rotationally
invariant dot products compare directional features already accumulated on each
atom, both the channel projections and angular contractions scale with the
number of atoms rather than edges and require no explicit Clebsch--Gordan tensor
products.

\noindent{\bfseries Geometric Tensor Injection.}
Geometric Tensor Injection constructs new directional features by coupling the
scalar intermediate to edge geometry. We first project this intermediate on each
atom as $\mathbf p_j^\ell=W_{\rm inj}^\ell\mathbf x_j$
and then modulate it on edges with the precomputed radial filters and spherical
harmonics:
\begin{equation}
  \mathbf J_{i,m}^\ell=\sum_{j\in\mathcal N(i)}
  \bigl(\mathbf p_j^\ell\odot\boldsymbol\xi_{ij}^\ell\bigr)
  Y_{\ell m}(\widehat{\mathbf r}_{ij}).
  \label{eq:facto-tensor-injection}
\end{equation}
Because the radial filter $\boldsymbol\xi_{ij}^\ell$ is precomputed once and
shared across blocks, no radial network is evaluated inside the interaction
block, and the remaining gather, elementwise product and neighbor sum fuse into
atom-centered aggregation.

\noindent{\bfseries Tensor Transport.}
Tensor Transport propagates existing angular information between neighboring
atoms. An edgewise formulation would compute invariant alignment scores by
contracting incoming tensors with edge spherical harmonics and then mixing
channels on every edge:
\begin{equation}
  \mathbf a_{ij}^\ell
    =\left[\sum_{m=-\ell}^{\ell}
      \mathbf T_{j,m}^\ell Y_{\ell m}(\widehat{\mathbf r}_{ij})\right]
      W_{\rm align}^\ell.
  \label{eq:facto-edgewise-alignment}
\end{equation}
Instead, exploiting linearity to avoid a dense projection on each edge, we
project each atom's tensor once and reuse it across all of its incident edges:
\begin{equation}
  \mathbf A_j^\ell=\mathbf T_j^\ell W_{\rm align}^\ell,
  \qquad
  \mathbf a_{ij}^\ell=\sum_{m=-\ell}^{\ell}
    \mathbf A_{j,m}^\ell Y_{\ell m}(\widehat{\mathbf r}_{ij}).
  \label{eq:facto-atomwise-alignment}
\end{equation}
We similarly precompute the projected tensor messages
$\mathbf M_j^\ell=\mathbf T_j^\ell W_{\rm msg}^\ell$ and scalar gates
$W_{\rm gate}^\ell\mathbf x_j$ on atoms before forming the gated neighbor sum:
\begin{equation}
  \begin{aligned}
    \mathbf g_{ij}^\ell
      &=\chi_{ij}(W_{\rm gate}^\ell\mathbf x_j)\odot
        \mathbf a_{ij}^\ell,\\
    \mathbf P_{i,m}^\ell
      &=\sum_{j\in\mathcal N(i)}
        \mathbf g_{ij}^\ell\odot\mathbf M_{j,m}^\ell.
  \end{aligned}
  \label{eq:facto-tensor-transport}
\end{equation}
Because the alignment gate $\mathbf g_{ij}^\ell$ is rotationally invariant, it
preserves the tensor transformation law of $\mathbf M_j^\ell$ while leaving only
angular contractions, elementwise products and neighbor reductions on edges.

\noindent{\bfseries Atom-centered aggregation.}
By eliminating dense linear operations from the three pairwise paths---Factorized
Scalar Attention, Geometric Tensor Injection and Tensor Transport---we can form
and accumulate each neighbor contribution directly at the receiving atom without
storing intermediate edge messages
(Fig.~\ref{fig:facto-architecture}c).

\noindent{\bfseries Residual updates.}
At the end of block $t$, we update the stored scalar and tensor states with a
fixed residual scale $\alpha$:
\begin{equation}
  \begin{aligned}
    \mathbf h_i^{(t+1)}
      &=\mathbf h_i^{(t)}+\alpha\Delta\mathbf h_i,\\
    \mathbf T_i^{(t+1,\ell)}
      &=\mathbf T_i^{(t,\ell)}+\alpha(\mathbf J_i^\ell+\mathbf P_i^\ell).
  \end{aligned}
  \label{eq:facto-residual-updates}
\end{equation}

\noindent{\bfseries Graph readout.}
After all $K$ interaction blocks, we concatenate the post-block scalar states as
$\mathbf h_i^{\rm cat}=[\mathbf h_i^{(1)};\ldots;\mathbf h_i^{(K)}]$.
Let $\mathbf h_Z^{\rm iso}$ denote the corresponding state for an isolated atom
of element $Z$, evaluated with the same parameters and no neighbors.
We apply a bias-free shared linear projection $\mathbf w$ to define per-atom energy contributions:
\begin{equation}
  E_i=e_{Z_i}^{\rm ref}+\delta_{Z_i}
    +\lambda_{Z_i}\mathbf w^\top
      (\mathbf h_i^{\rm cat}-\mathbf h_{Z_i}^{\rm iso}),
  \label{eq:facto-atomic-readout}
\end{equation}
where $e_Z^{\rm ref}$ is a fixed elemental reference, $\lambda_Z$ is a learned
elemental scale and $\delta_Z$ is a learned elemental shift initialized to zero.
The isolated-atom energy is therefore $e_Z^{\rm ref}+\delta_Z$; geometry-independent
reference terms do not contribute to forces or stress.
We obtain the total potential energy by summing
these per-atom contributions together with a short-range screened pair-repulsion
term $E_{\rm rep}$\cite{ziegler1985stopping} inspired by the implementation in MatGL\cite{Ko2025}:
\begin{equation}
  E=\sum_i E_i+E_{\rm rep},
  \label{eq:facto-total-energy}
\end{equation}
from which forces and stress follow as analytical derivatives.

\subsection{Materials discovery and inference scaling}
\label{sec:matbench-results}

We first evaluate FACTO on Matbench Discovery\cite{riebesell2025matbench,matbench2026snapshot},
which jointly assesses crystal stability and thermal conductivity (CPS),
finite-temperature dynamical fidelity (CMDS)\cite{gawkowski2026dynamat} and
short-range pairwise physicality (CDS). Across the 46 ranked models, comprising the 42
public default-cohort models with complete score triplets plus the four
FACTO-OAM variants, the FACTO-OAM family occupies the top four overall positions
when ranked by the sum of CPS, CMDS and CDS ranks, ahead of TACE-OAM-L
and ORB~v3 (Fig.~\ref{fig:facto-overview}a and Table~\ref{tab:facto-discovery};
component metrics in Appendix Note~3). Whereas models with slightly higher
static CPS, such as Prophet-OAME-MBD and EquiformerV3+DeNS-OAM, rank 26 and 24
overall because of lower dynamical and diatomic scores, the FACTO-OAM family
maintains high accuracy across all three axes: FACTO-OAM-XL ranks in the top
eight on every composite metric, while FACTO-OAM-S and FACTO-OAM-M take the top
two positions in CMDS and ranks 3 and 4 in CDS. This accuracy is
paired with an order-of-magnitude advantage in inference throughput
(Fig.~\ref{fig:facto-overview}b,c). Along with
Prophet-OAME-MBD, the four FACTO-OAM variants define the CPS--throughput Pareto
frontier on 1,000-atom silicon (Fig.~\ref{fig:facto-overview}b). At the
high-accuracy end, FACTO-OAM-XL is more than $15\times$ faster than
Prophet-OAME-MBD, EquiformerV3+DeNS-OAM and eSEN-30M-OAM, and roughly
$3\text{--}4\times$ faster than GRACE-3L-OAM-L and TACE-OAM-L at comparable CPS.
Across the faster tiers, FACTO-OAM-L through FACTO-OAM-S achieve throughputs of
64.88 to 149.6~call/s, outpacing lighter baselines including NequIP-OAM-L,
GRACE-2L-OAM-L, ORB~v3 and MACE-MPA-0 while achieving superior overall rank
sums (with FACTO-OAM-L also surpassing all four in CPS and FACTO-OAM-M and
FACTO-OAM-S trading a small CPS reduction for higher dynamical scores and speed).
This efficiency stems from eliminating edge matrix multiplications and
persistent edge state rather than from static graph pruning. At larger system
sizes, we further accelerate inference with custom aggregation and tensor-core
emulation kernels (Section~\ref{sec:facto-acceleration}). Beyond the smallest
8-atom cell, FACTO maintains lower latency across all evaluated system sizes
(Fig.~\ref{fig:facto-overview}c).

\begin{table}[!t]
\centering
\caption{{\bfseries Matbench Discovery scores, ranks and inference throughput.}
Dataset codes denote training data: O = OMat24\cite{barrosluque2026omat}, M = MPtrj\cite{deng2023chgnet}, A = Alexandria\cite{schmidt2024alexandria} and E = ELEMENTA\cite{kairos2026elementa} (MPA = MPtrj + Alexandria). For the 46 ranked models (four FACTO-OAM variants and 42 public models with complete scores; Fig.~\ref{fig:facto-overview}a)\cite{matbench2026snapshot}, CPS, CMDS, CDS and Sum-of-ranks are reported as value/rank; the FACTO-MPA series is listed below without ranks. Throughput is measured on 1,000- and 8,000-atom silicon supercells (Fig.~\ref{fig:facto-overview}b); OOM denotes out of memory. Bold values mark the best ranked and overall results in each column. Component metrics and the full public cohort are reported in Appendix Note~3.}
\label{tab:facto-discovery}
\begingroup
\small
\setlength{\tabcolsep}{4pt}
\renewcommand{\arraystretch}{1.15}
\newcommand{\TableHeader}[1]{\multirow{2}{*}[-\dimexpr(\aboverulesep+\belowrulesep+\cmidrulewidth)/2\relax]{#1}}
\begin{tabular*}{\linewidth}{@{\extracolsep{\fill}}llrrrrrr@{}}
\toprule
\TableHeader{Model} & \TableHeader{Dataset} & \TableHeader{CPS $\uparrow$} & \TableHeader{CMDS $\uparrow$} & \TableHeader{CDS $\uparrow$} & \TableHeader{Sum-of-ranks $\downarrow$} & \multicolumn{2}{c@{}}{Throughput $\uparrow$ (call/s)} \\
\cmidrule(l){7-8}
 & & & & & & 1,000 Si & 8,000 Si \\
\midrule
\textcolor[HTML]{133F72}{{\bfseries FACTO-OAM-XL}} & OAM & 0.890\makebox[1.25em][l]{{\scriptsize /7}} & 0.751\makebox[1.25em][l]{{\scriptsize /6}} & 0.771\makebox[1.25em][l]{{\scriptsize /8}} & {\bfseries 21\makebox[1.25em][l]{{\scriptsize /1}}} & 29.84 & 6.858 \\
\textcolor[HTML]{3567A3}{{\bfseries FACTO-OAM-L}} & OAM & 0.883\makebox[1.25em][l]{{\scriptsize /12}} & 0.773\makebox[1.25em][l]{{\scriptsize /3}} & 0.781\makebox[1.25em][l]{{\scriptsize /7}} & 22\makebox[1.25em][l]{{\scriptsize /2}} & 64.88 & 14.30 \\
\textcolor[HTML]{0072B2}{{\bfseries FACTO-OAM-M}} & OAM & 0.863\makebox[1.25em][l]{{\scriptsize /17}} & 0.789\makebox[1.25em][l]{{\scriptsize /2}} & 0.822\makebox[1.25em][l]{{\scriptsize /4}} & 23\makebox[1.25em][l]{{\scriptsize /3}} & 117.3 & 39.33 \\
\textcolor[HTML]{00839B}{{\bfseries FACTO-OAM-S}} & OAM & 0.855\makebox[1.25em][l]{{\scriptsize /19}} & {\bfseries 0.792\makebox[1.25em][l]{{\scriptsize /1}}} & 0.825\makebox[1.25em][l]{{\scriptsize /3}} & 23\makebox[1.25em][l]{{\scriptsize /3}} & {\bfseries 149.6} & {\bfseries 53.69} \\
\midrule
Prophet-OAME-MBD\cite{kairos2026prophet} & OAME & {\bfseries 0.912\makebox[1.25em][l]{{\scriptsize /1}}} & 0.578\makebox[1.25em][l]{{\scriptsize /43}} & 0.572\makebox[1.25em][l]{{\scriptsize /28}} & 72\makebox[1.25em][l]{{\scriptsize /26}} & 1.924 & OOM \\
EquiformerV3+DeNS-OAM\cite{liao2026equiformerv3} & OAM & 0.902\makebox[1.25em][l]{{\scriptsize /4}} & 0.638\makebox[1.25em][l]{{\scriptsize /24}} & 0.269\makebox[1.25em][l]{{\scriptsize /43}} & 71\makebox[1.25em][l]{{\scriptsize /24}} & 1.833 & OOM \\
GRACE-3L-OAM-L\cite{lysogorskiy2025grace} & OAM & 0.900\makebox[1.25em][l]{{\scriptsize /5}} & 0.739\makebox[1.25em][l]{{\scriptsize /9}} & 0.699\makebox[1.25em][l]{{\scriptsize /17}} & 31\makebox[1.25em][l]{{\scriptsize /7}} & 10.08 & 1.308 \\
TACE-OAM-L\cite{xu2025tace} & OAM & 0.889\makebox[1.25em][l]{{\scriptsize /8}} & 0.728\makebox[1.25em][l]{{\scriptsize /11}} & 0.761\makebox[1.25em][l]{{\scriptsize /9}} & 28\makebox[1.25em][l]{{\scriptsize /5}} & 6.878 & 1.085 \\
eSEN-30M-OAM\cite{fu2025esen} & OAM & 0.888\makebox[1.25em][l]{{\scriptsize /9}} & 0.625\makebox[1.25em][l]{{\scriptsize /30}} & 0.648\makebox[1.25em][l]{{\scriptsize /24}} & 63\makebox[1.25em][l]{{\scriptsize /18}} & 1.779 & OOM \\
NequIP-OAM-XL\cite{tan2025nequip} & OAM & 0.886\makebox[1.25em][l]{{\scriptsize /11}} & 0.623\makebox[1.25em][l]{{\scriptsize /31}} & 0.796\makebox[1.25em][l]{{\scriptsize /6}} & 48\makebox[1.25em][l]{{\scriptsize /10}} & 15.67 & 2.171 \\
NequIP-OAM-L\cite{tan2025nequip} & OAM & 0.870\makebox[1.25em][l]{{\scriptsize /15}} & 0.670\makebox[1.25em][l]{{\scriptsize /15}} & 0.818\makebox[1.25em][l]{{\scriptsize /5}} & 35\makebox[1.25em][l]{{\scriptsize /8}} & 42.11 & 6.902 \\
GRACE-2L-OAM-L\cite{lysogorskiy2025grace} & OAM & 0.865\makebox[1.25em][l]{{\scriptsize /16}} & 0.759\makebox[1.25em][l]{{\scriptsize /4}} & 0.694\makebox[1.25em][l]{{\scriptsize /19}} & 39\makebox[1.25em][l]{{\scriptsize /9}} & 20.66 & 2.923 \\
ORB v3\cite{rhodes2025orb3} & OAM & 0.860\makebox[1.25em][l]{{\scriptsize /18}} & 0.736\makebox[1.25em][l]{{\scriptsize /10}} & {\bfseries 0.834\makebox[1.25em][l]{{\scriptsize /1}}} & 29\makebox[1.25em][l]{{\scriptsize /6}} & 14.29 & 1.937 \\
MACE-MPA-0\cite{batatia2025foundation} & MPA & 0.795\makebox[1.25em][l]{{\scriptsize /26}} & 0.659\makebox[1.25em][l]{{\scriptsize /16}} & 0.728\makebox[1.25em][l]{{\scriptsize /13}} & 55\makebox[1.25em][l]{{\scriptsize /12}} & 19.40 & 5.229 \\
\midrule
\textcolor[HTML]{133F72}{{\bfseries FACTO-MPA-XL}} & MPA & 0.867\phantom{\makebox[1.25em][l]{}} & 0.756\phantom{\makebox[1.25em][l]{}} & 0.770\phantom{\makebox[1.25em][l]{}} & \textemdash\phantom{\makebox[1.25em][l]{}} & 29.84 & 6.858 \\
\textcolor[HTML]{3567A3}{{\bfseries FACTO-MPA-L}} & MPA & 0.859\phantom{\makebox[1.25em][l]{}} & 0.775\phantom{\makebox[1.25em][l]{}} & 0.801\phantom{\makebox[1.25em][l]{}} & \textemdash\phantom{\makebox[1.25em][l]{}} & 64.88 & 14.30 \\
\textcolor[HTML]{0072B2}{{\bfseries FACTO-MPA-M}} & MPA & 0.847\phantom{\makebox[1.25em][l]{}} & {\bfseries 0.793\phantom{\makebox[1.25em][l]{}}} & 0.813\phantom{\makebox[1.25em][l]{}} & \textemdash\phantom{\makebox[1.25em][l]{}} & 117.3 & 39.33 \\
\textcolor[HTML]{00839B}{{\bfseries FACTO-MPA-S}} & MPA & 0.832\phantom{\makebox[1.25em][l]{}} & 0.790\phantom{\makebox[1.25em][l]{}} & 0.807\phantom{\makebox[1.25em][l]{}} & \textemdash\phantom{\makebox[1.25em][l]{}} & {\bfseries 149.6} & {\bfseries 53.69} \\
\bottomrule
\end{tabular*}
\endgroup
\end{table}

Trained from scratch on MPtrj and subsampled Alexandria (sAlex) without OMat24
pretraining, the FACTO-MPA series also achieves strong performance across all
three benchmarks at the same high throughput (Table~\ref{tab:facto-discovery},
bottom block). On the same training corpus, both FACTO-MPA-XL and FACTO-MPA-L
substantially outperform MACE-MPA-0 across CPS, CMDS, CDS and speed, and match
or surpass OAM-pretrained baselines such as GRACE-2L-OAM-L and ORB~v3 in CPS,
while FACTO-MPA-M and FACTO-MPA-S closely match their OAM counterparts on CMDS
and CDS. Comparing the two families indicates that, with MPA training alone,
the models produce smooth molecular-dynamics trajectories and diatomic curves,
whereas OMat24 pretraining primarily improves equilibrium stability and anharmonic phonon
transport (reducing $\kappa_{\rm SRME}$ from 0.2218 in FACTO-MPA-XL to 0.1560 in
FACTO-OAM-XL; Appendix Table~\ref{tab:si-mbd-core}).

Within both families, scaling from L to XL brings a modest gain in CPS while
slightly lowering CMDS and CDS, reflecting the runtime subscore penalty at
larger model sizes and higher diatomic force-reversal counts on steep repulsive
walls (Appendix Tables~\ref{tab:si-mbd-md}
and~\ref{tab:si-mbd-diatomic-physicality}). This plateau reflects both
diminishing architectural returns once Tensor Transport and $\ell\leq4$
equivariant channels are active in FACTO-L and our use of a uniform
training recipe across all four model sizes without size-specific
hyperparameter or loss-weight tuning (Methods and Appendix Note~1). Achieving
rank-1 overall standing and leading throughput under this fixed recipe
underscores that FACTO's performance stems from its atom-centered equivariant
design while leaving scope for training-recipe optimization at the largest scale
(Discussion).
\FloatBarrier

\Needspace{5\baselineskip}
\subsection{Out-of-equilibrium stability and reactive dynamics}
\label{sec:arena-results}

To probe physical reliability far from equilibrium, we evaluate the FACTO-OAM
family on MLIP Arena\cite{chiang2025arena,arena2026snapshot}, where it achieves
state-of-the-art performance across pairwise potentials, bulk deformation scans,
extreme thermodynamic stress tests and reactive molecular dynamics
(Fig.~\ref{fig:facto-arena}; full diagnostics in Appendix Note~3). On static
diatomic scans (Fig.~\ref{fig:facto-arena}a), all four FACTO-OAM variants
have low finite-grid force--energy conservation deviations of
$0.01360\text{--}0.01707$~eV\,\AA$^{-1}$,
more than $4\times$ smaller than those of MACE-MPA and
MACE-MP(M)\cite{batatia2025foundation} and orders of magnitude smaller than those of
ORBv2\cite{neumann2024orb},
CHGNet\cite{deng2023chgnet} and SevenNet\cite{park2024sevennet} (Appendix
Table~\ref{tab:si-arena-diatomics}). For volumetric crystal deformations across
all 1,000 WBM structures\cite{wang2021wbm} (Fig.~\ref{fig:facto-arena}b), both
relaxed equation-of-state and fixed-coordinate energy--volume tortuosities
remain near unity ($1.000\text{--}1.011$) with $100\%$ completion (Appendix
Tables~\ref{tab:si-arena-eos_bulk} and~\ref{tab:si-arena-ev}).

\setlength{\textfloatsep}{12pt plus 2pt minus 2pt}
\begin{figure}[!t]
\centering
\includegraphics[width=\linewidth]{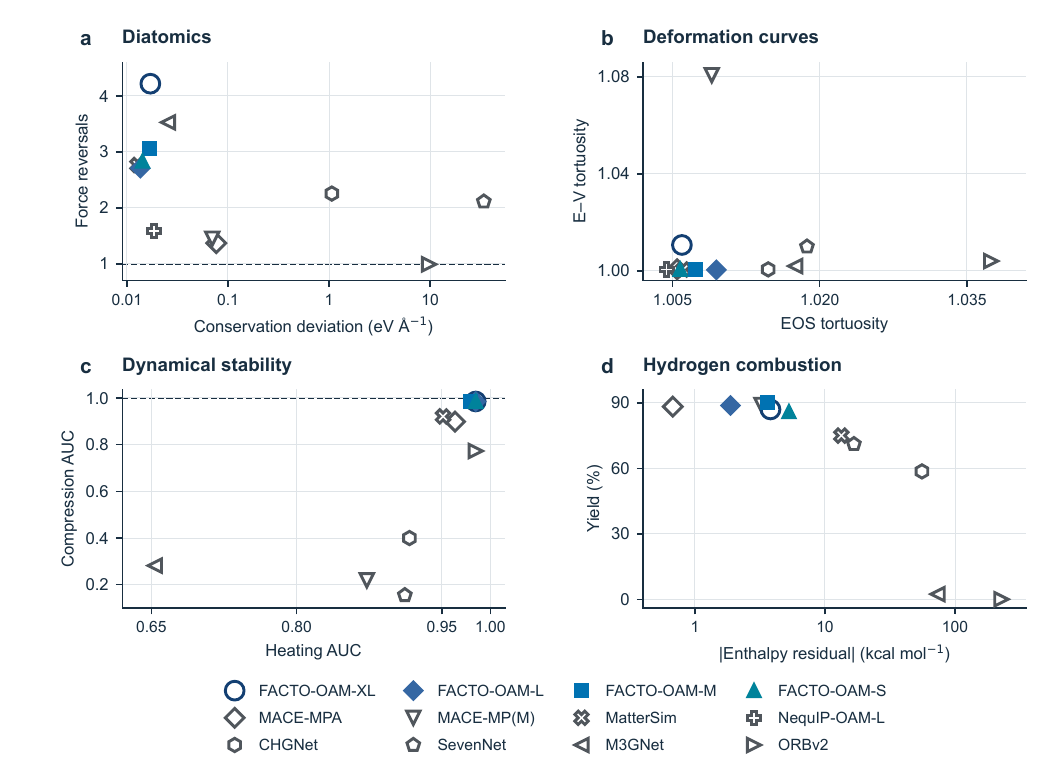}
\caption{{\bfseries Out-of-equilibrium physicality, dynamical stability and reactive dynamics on MLIP Arena.}
{\bfseries a,} Mean diatomic force reversals versus finite-grid force--energy
conservation deviation on a logarithmic x-axis; the dashed line marks the ideal
reversal count of one.
{\bfseries b,} Tortuosity for relaxed
equation-of-state (EOS) and fixed-coordinate energy--volume (E--V) curves.
Both axes are linear; values closer to one indicate smoother curves.
{\bfseries c,} Area under the heating and compression survival curves (AUC);
dashed lines mark the discrete 100-bin scorer ceiling of $65/66\approx0.9848$ (continuous ideal of one).
Coincident FACTO points (FACTO-OAM-XL, FACTO-OAM-L and FACTO-OAM-S at $(0.9848, 0.9848)$) are shown
with nested symbols at their measured coordinates.
{\bfseries d,} Water yield
versus absolute potential-energy reaction-enthalpy residual in hydrogen
combustion on a logarithmic x-axis.
For CHGNet and M3GNet, the published combustion endpoints are from partial trajectories.
Public baseline values are from the MLIP Arena leaderboard\cite{arena2026snapshot};
full FACTO-OAM diagnostics are reported in Appendix Note~3.}
\label{fig:facto-arena}
\end{figure}

This smooth potential energy surface translates directly into trajectory
stability in the molecular-dynamics stress benchmark
(Fig.~\ref{fig:facto-arena}c), which measures the area under the finite-energy
survival curve (AUC) during heating ramps from 300 to 3,000~K and compression
ramps up to 100~GPa. Most public baselines exhibit substantially reduced trajectory survival
under high-pressure compression: SevenNet, MACE-MP(M),
M3GNet\cite{chen2022m3gnet} and CHGNet fall to compression AUCs of
$0.1550\text{--}0.3998$, ORBv2 drops from 0.9848 in heating to 0.7725 in
compression, and MACE-MPA (0.9635 and 0.8986) and
MatterSim\cite{yang2024mattersim} (0.9511 and 0.9207) also fall short of the
benchmark's discrete 100-bin ceiling ($65/66\approx0.9848$). By contrast, all four
FACTO-OAM variants cluster at the upper-right of the plot, with FACTO-OAM-XL,
FACTO-OAM-L and FACTO-OAM-S reaching the 0.9848 ceiling in both heating and
compression AUC and FACTO-OAM-M recording 0.9799 and 0.9848, surpassing every
compared baseline in joint thermal and barostat robustness across all 400
trajectories (Appendix Table~\ref{tab:si-arena-stability}).

This dynamical stability extends to the 1~ns (2,000,000-step) reactive hydrogen
combustion benchmark on 384 atoms ($128\,\mathrm{H_2}+64\,\mathrm{O_2}$;
Fig.~\ref{fig:facto-arena}d). Among the public baselines, trajectories evaluated with CHGNet and M3GNet
terminate prematurely ($26.30\%$ and $22.70\%$ completion) with large enthalpy
residuals ($>55$~kcal\,mol$^{-1}$), whereas the ORBv2 trajectory reaches completion without ignition
(0\% water yield and a 227.5~kcal\,mol$^{-1}$ residual). With MatterSim and
SevenNet, water yields are moderate ($75.00\%$ and $71.09\%$), with residuals of
13.38 and 16.70~kcal\,mol$^{-1}$. By contrast, for all four FACTO-OAM variants,
the reactive trajectories reach completion in the high-yield, low-residual regime
alongside MACE-MPA and MACE-MP(M), achieving $85.94\text{--}90.10\%$ water
yield, with FACTO-OAM-M reaching the highest yield ($90.10\%$) among all
compared models and FACTO-OAM-L reaching $88.80\%$. Across the series, we also
observe small reaction-enthalpy residuals ($1.878\text{--}5.288$~kcal\,mol$^{-1}$,
down to 1.878~kcal\,mol$^{-1}$ for FACTO-OAM-L) and negligible center-of-mass
drift ($0.001247\text{--}0.001764$~\AA), while sustaining rollout throughputs of
$45.41\text{--}157.7$~steps\,s$^{-1}$ (Appendix
Table~\ref{tab:si-arena-combustion}).

\Needspace{10\baselineskip}
\subsection{Crystalline elastic, thermal and force response}
\label{sec:matcalc-results}

We next evaluate crystalline property prediction on
MatCalc\cite{kaplan2025matpes,matcalc2026software,deng2025softening}, comparing
the FACTO-OAM series against five published baselines (NequIP-OAM-XL and
NequIP-OAM-L\cite{batzner2022nequip,tan2025nequip},
Allegro-OAM-L\cite{musaelian2023allegro,tan2025nequip},
MatterSim-5M\cite{yang2024mattersim} and
\mbox{UMA-s}\cite{kavanagh2026foundation}) across equilibrium structure,
energetics and continuum elasticity (Fig.~\ref{fig:facto-matcalc}a--d; full
values in Appendix Note~3). Despite operating at substantially higher inference
throughput than competing equivariant models (Table~\ref{tab:facto-discovery}),
the FACTO-OAM family matches or surpasses the baselines across these
observables. All four FACTO-OAM variants have lower errors than every published baseline
in relaxed CrystalNN structural-fingerprint distance ($0.1959\text{--}0.2291$
versus $0.2670\text{--}0.3220$; Fig.~\ref{fig:facto-matcalc}a) as well as
Voigt--Reuss--Hill bulk-modulus ($11.05\text{--}12.58$ versus
$12.70\text{--}16.90$~GPa; Fig.~\ref{fig:facto-matcalc}c) and shear-modulus
($12.04\text{--}13.54$ versus $14.10\text{--}21.50$~GPa, led by FACTO-OAM-L at
12.04~GPa; Fig.~\ref{fig:facto-matcalc}d) MAEs. On formation energy
(Fig.~\ref{fig:facto-matcalc}b), FACTO-OAM-XL achieves the lowest error among
all compared models (22.35~meV\,atom$^{-1}$), while FACTO-OAM-M, FACTO-OAM-L
and FACTO-OAM-S ($24.73\text{--}26.06$~meV\,atom$^{-1}$) perform on par with
Allegro-OAM-L, UMA-s, NequIP-OAM-XL and NequIP-OAM-L
($24.70\text{--}26.60$~meV\,atom$^{-1}$) and substantially ahead of
MatterSim-5M (40.50~meV\,atom$^{-1}$).

\begin{figure}[!t]
\centering
\includegraphics[width=\linewidth]{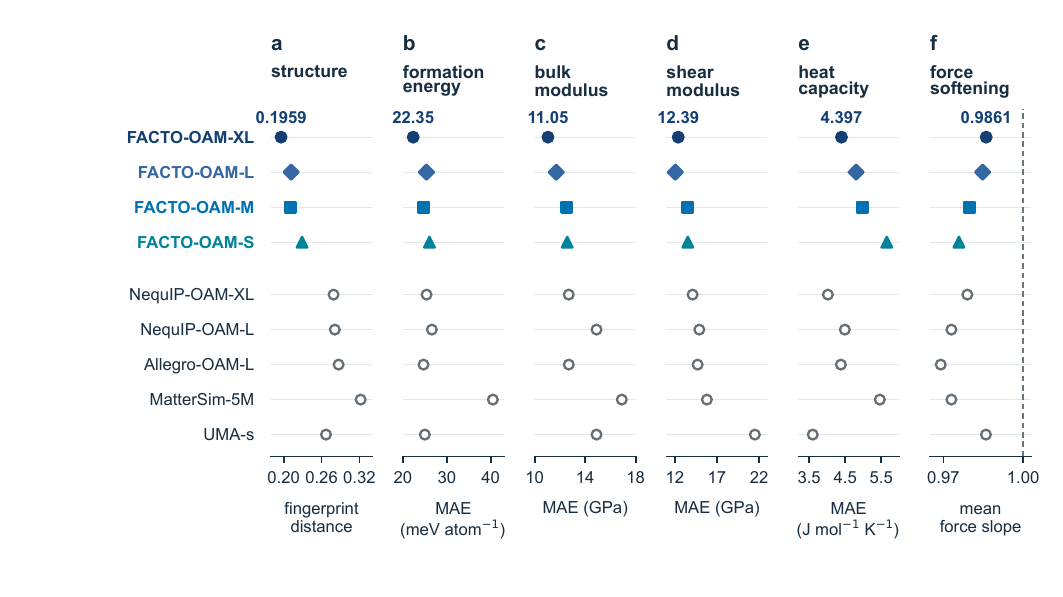}
\caption{{\bfseries Crystalline-property accuracy and off-equilibrium force amplitudes.}
{\bfseries a--e,} CrystalNN structural-fingerprint distance, formation-energy
MAE, bulk- and shear-modulus MAEs, and 300~K harmonic heat-capacity MAE;
lower values indicate smaller errors.
{\bfseries f,} Mean proportional slope between predicted and DFT force
components; the dashed line marks the ideal value of one.
Filled circles, diamonds, squares and triangles denote FACTO-OAM-XL,
FACTO-OAM-L, FACTO-OAM-M and FACTO-OAM-S, respectively; open circles denote published baselines.
Top annotations give the FACTO-OAM-XL values. Each panel uses a separately labeled linear scale.
Published baseline values are taken from ref.~\citenum{kavanagh2026foundation}.
Complete property values are reported in Appendix Note~3.}
\label{fig:facto-matcalc}
\end{figure}

Comparable accuracy extends to vibrational thermodynamics and off-equilibrium
force calibration (Fig.~\ref{fig:facto-matcalc}e,f). For 300~K harmonic heat
capacity across 1,170 crystals (Fig.~\ref{fig:facto-matcalc}e), where
finite-displacement force constants are sensitive to small fluctuations in predicted forces,
FACTO-OAM-XL (4.397~J\,mol$^{-1}$\,K$^{-1}$) and FACTO-OAM-L
(4.801~J\,mol$^{-1}$\,K$^{-1}$) perform comparably to Allegro-OAM-L
(4.380~J\,mol$^{-1}$\,K$^{-1}$) and NequIP-OAM-L
(4.490~J\,mol$^{-1}$\,K$^{-1}$), while trailing UMA-s
(3.600~J\,mol$^{-1}$\,K$^{-1}$) and NequIP-OAM-XL
(4.020~J\,mol$^{-1}$\,K$^{-1}$). Conversely, on high-energy WBM configurations
(Fig.~\ref{fig:facto-matcalc}f), FACTO-OAM-XL (0.9861) and FACTO-OAM-L (0.9847)
closely approach the ideal force-calibration slope of unity, nearly eliminating
systematic force softening. Even the fastest variants, FACTO-OAM-S and
FACTO-OAM-M, which run $2.785\text{--}9.546\times$ faster than NequIP-OAM-L and
NequIP-OAM-XL on 1,000-atom silicon (Table~\ref{tab:facto-discovery}),
outperform both NequIP-OAM models on structural relaxation and elastic moduli
while matching their formation-energy and force-slope accuracy.
\FloatBarrier

\Needspace{10\baselineskip}
\subsection{Accelerating FACTO inference}
\label{sec:facto-acceleration}

Because FACTO separates matrix-multiplication-free neighbor reductions from
dense atom-wise channel mixing, its inference can be accelerated through two
orthogonal optimizations in addition to PyTorch ahead-of-time Inductor (AOTI)
compilation\cite{ansel2024pytorch2}. First, although pointwise edge operations are
fused across the unified energy, force and stress graph in AOTI, standard scatter
reductions still incur global-memory traffic and contention from GPU atomic additions.
Our custom atom-centered forward kernels avoid materializing intermediate edge
messages and eliminate GPU atomic additions to the aggregation outputs:
neighbor contributions are accumulated in registers, and each output component
is written once with an ordinary store
(Fig.~\ref{fig:facto-architecture}c). Neighbor indices, shared geometry and final
outputs still require global-memory accesses. Second, to accelerate the dense atom-wise
linear projections in both the forward energy evaluation and backward differentiation
for forces and stress while retaining near-single-precision numerical fidelity, we apply
tensor-core FP32 emulation via a three-product bfloat16 decomposition
(BF16$\times$3) that recovers the leading FP32 residual
terms\cite{henry2019bfloat16} (Appendix Note~2).

\begin{figure}[!t]
\centering
\includegraphics[width=\linewidth]{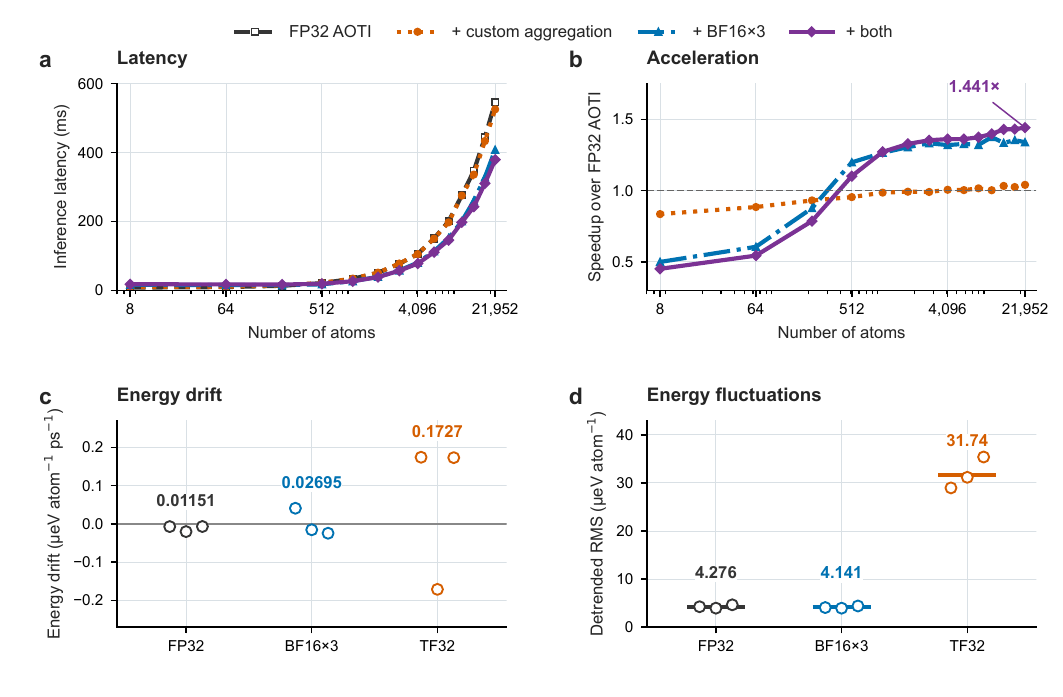}
\caption{{\bfseries Faster large-system evaluation with controlled energy fluctuations.}
{\bfseries a,} FACTO-XL inference latency for diamond-silicon supercells.
{\bfseries b,} Speedup relative to plain FP32 AOTI.
The combination of custom aggregation and FP32 emulation (labeled BF16$\times$3)
is compared with each optimization alone; the annotation marks the peak combined
speedup ($1.441\times$ at 21,952 atoms).
{\bfseries c,d,} Signed total-energy drift and detrended RMS fluctuation
for three 500~ps trajectories per arithmetic mode using FACTO-OAM-M, 64-atom
silicon and a 1~fs timestep. Each point represents an independent trajectory.
Numbers in c give the mean absolute drift across the three
trajectories for each arithmetic mode. Short horizontal bars and numbers in d show the geometric means.
Under FP32 emulation (BF16$\times$3), fluctuation amplitudes remain close to those
of FP32, whereas they are substantially increased under TF32.
Total-energy traces are shown in Appendix Fig.~\ref{fig:nve-traces}.}
\label{fig:facto-acceleration}
\end{figure}

Targeting memory-bound neighbor aggregation and compute-bound node projections
together yields compounding speedups as the simulated system grows
(Fig.~\ref{fig:facto-acceleration}a,b). On the smallest diamond-silicon
supercells ($N\leq216$ atoms), kernel-launch and operand-splitting overheads
outweigh the savings from custom aggregation and BF16$\times$3. Once the
workload saturates the GPU ($N\geq1,000$ atoms), combining both optimizations
achieves the lowest latency at every system size, with custom aggregation also
reducing latency on its own for $N\geq4,096$ atoms. For the wide
($D=2,048$, $L=6$) FACTO-XL backbone, where dense atom-wise projections dominate
runtime, BF16$\times$3 alone delivers a $1.339\times$ speedup at 21,952 atoms,
and adding custom receiver accumulation increases the combined speedup to
$1.441\times$ (reducing calculator-call latency from 546.3 to 379.1~ms).

Tensor-core FP32 emulation avoids the dynamical degradation caused by
standard reduced-precision tensor-core arithmetic
(Fig.~\ref{fig:facto-acceleration}c,d). Across three 500~ps microcanonical
(NVE) simulations of 64-atom silicon with FACTO-OAM-M ($\Delta t=1$~fs),
standard TF32 arithmetic inflates the geometric-mean detrended total-energy
fluctuation by $7.424\times$ relative to strict FP32 (31.74 versus
4.276~$\mu$eV\,atom$^{-1}$) and increases the mean absolute energy drift by
$15.00\times$ (0.1727 versus 0.01151~$\mu$eV\,atom$^{-1}$\,ps$^{-1}$; Appendix
Table~\ref{tab:nve-all-seeds} and Appendix Fig.~\ref{fig:nve-traces}). By
recovering the leading residual bits of each matrix product, BF16$\times$3
restores the geometric-mean detrended energy fluctuation to the strict-FP32
level (4.141 versus 4.276~$\mu$eV\,atom$^{-1}$) while keeping the mean absolute
drift in the $10^{-2}$~$\mu$eV\,atom$^{-1}$\,ps$^{-1}$ regime
(0.02695~$\mu$eV\,atom$^{-1}$\,ps$^{-1}$). Tensor-core FP32 emulation thus
provides a transferable path to accelerate dense equivariant projections while
preserving the energy conservation required for long-horizon molecular dynamics.

\FloatBarrier
\section{Discussion}

We introduced FACTO, an equivariant interatomic potential that
confines evolving learned state and dense channel mixing to atoms while
factoring pairwise interactions into lightweight, fully fusible neighbor
reductions.
By pairing this atom-centered architecture with custom register-accumulated
neighbor kernels and three-product bfloat16 tensor-core FP32 emulation
(BF16$\times$3), we accelerate both memory-bound edge aggregation and
compute-bound node projections while retaining strict-FP32-level energy
fluctuations under BF16$\times$3 in microcanonical dynamics.
Across Matbench Discovery, MLIP Arena and MatCalc, the FACTO family advances the
accuracy--throughput frontier, combining leading positions in the overall
Matbench Discovery rank-sum comparison and state-of-the-art dynamical and
reactive stability on MLIP Arena with competitive crystalline-property
predictions at substantially higher throughput than existing equivariant
potentials.

Several limitations suggest directions for future work.
First, wide atom-wise projections and high angular orders ($D$ up to 2,048 and
$L$ up to 6) increase parameter counts and per-atom memory footprints in our
largest models, restricting single-GPU evaluation of FACTO-L and FACTO-XL to
smaller maximum atom counts than FACTO-S and FACTO-M
(Fig.~\ref{fig:facto-overview}c).
Parameter sharing, low-rank node projections or multi-GPU domain decomposition
will be needed to extend the largest backbones to larger systems.
Second, computational constraints limited training-recipe exploration.
We used the same optimizer hyperparameters across model scales, concatenated
sAlex and MPtrj without dataset reweighting, adopted a two-stage conservative
OAM pipeline without auxiliary denoising objectives, and fixed the energy,
force and stress loss weights at $1:1:1$.
Systematic recipe exploration should further improve accuracy at a fixed
inference cost.
Third, scaling from FACTO-L to FACTO-XL yields diminishing returns: modest
gains in static discovery, formation energy and bulk modulus alongside a slight
increase in shear-modulus error at higher latency.
Distinguishing training-recipe saturation from the intrinsic limits of
semi-local message passing remains an open challenge for foundation
interatomic potentials.

Looking forward, co-designing equivariant architectures, arithmetic schemes
and hardware-level execution opens promising avenues for atomistic modeling at
unprecedented scales.
By eliminating the memory bottlenecks of pairwise edge tensors and preserving
numerical fidelity under hardware-accelerated emulation, the
atom-centered paradigm provides a natural foundation for multi-GPU spatial
domain decomposition, making million-atom simulations and long-timescale
investigations of defect kinetics, crack propagation and phase transitions
accessible with quantum accuracy.
Beyond crystalline materials, extending these efficient equivariant operators
to multiphase interfaces, amorphous electrolytes and complex chemical reaction
networks could substantially broaden their physical domain.
Ultimately, pairing high-throughput foundational potentials with autonomous
active-learning workflows and robotic synthesis offers a practical path toward
accelerating scientific discovery across diverse materials spaces.

\Needspace{8\baselineskip}
\section{Methods}

\subsection{Model specification}
\label{sec:methods-model}

We implement FACTO using the model settings in Appendix Table~\ref{tab:methods-model-settings}.
Tensor Transport is omitted in the small and medium backbones (FACTO-S and FACTO-M) and enabled in the large and extra-large backbones (FACTO-L and FACTO-XL), whereas Geometric Tensor Injection and Tensor-to-scalar Feedback are retained in all four.
Only scalar features are updated in the final interaction block.
To construct periodic neighbor graphs, we include periodic images of atoms within 6~\AA\ and exclude zero-distance self-interactions.
The neighbor limit is given in Appendix Table~\ref{tab:methods-model-settings}.

We initialize scalar features by applying an affine projection and SiLU to fixed elemental descriptors.
For the Gaussian radial basis, we use centers
$\mu_n=(n-1)/(N_r-1)$ for $n=1,\ldots,N_r$, with
$N_r=32$ for FACTO-S and $N_r=64$ for FACTO-M, FACTO-L and FACTO-XL.
Real spherical harmonics are normalized according to the integral convention $\int_{S^2}|Y_{\ell m}|^2\,d\Omega=1$.

\subsection{Training data and optimization}
\label{sec:methods-training}

We train FACTO-MPA from random initialization on the MPtrj dataset released with CHGNet\cite{deng2023chgnet} and subsampled Alexandria (sAlex).
For FACTO-OAM, we first pretrain the model on OMat24 and then finetune it on the MPtrj--sAlex mixture\cite{barrosluque2026omat,schmidt2024alexandria}.
During finetuning, we retain the pretrained architecture and matching parameters but restart the optimizer and learning-rate schedule.
We concatenate MPtrj and sAlex without additional dataset balancing.

Fitting and validation partitions are kept separate during OMat24 pretraining.
In MPA training and OAM finetuning, the validation partitions are included in the training set and used to monitor training loss.
We use the final exponential-moving-average (EMA) checkpoint for evaluation.
Structures are randomly permuted and packed by atom count up to a per-device
limit; larger structures are omitted. Appendix Note~1 specifies
partitions, batching and stage hyperparameters.

Fixed elemental references are subtracted before evaluating the energy loss
and restored in physical predictions. Isolated-atom references are used in MPA
training and OAM finetuning, whereas OMat24 references are used in pretraining.
For $B$ structures with atom counts $N_b$, let $\mathcal A$ denote the set of atoms not marked as fixed. With prediction-minus-target differences denoted by
$\Delta$, the objective is
\begin{equation}
  \begin{aligned}
    \mathcal L&=w_E\mathcal L_E+w_F\mathcal L_F
      +w_\sigma\mathcal L_\sigma,\\
    \mathcal L_E&=\frac{1}{B}\sum_{b=1}^{B}
      \frac{|\Delta E_b|}{N_b},\qquad
    \mathcal L_F=\frac{1}{|\mathcal A|}
      \sum_{i\in\mathcal A}\|\Delta\mathbf F_i\|_2,\\
    \mathcal L_\sigma&=\frac{1}{9B}
      \sum_{b=1}^{B}\sum_{u,v=1}^{3}|\Delta\sigma_{b,uv}|.
  \end{aligned}
  \label{eq:methods-training-loss}
\end{equation}
We evaluate the loss terms in eV for energy, eV\,\AA$^{-1}$ for force and eV\,\AA$^{-3}$ for stress.
We average absolute energy errors per atom across structures, force-vector error norms across free atoms and absolute errors across all nine stress components.
In distributed training, loss reductions are normalized by the global counts of structures, free atoms and stress components.
Loss weights and shared optimizer settings are given in Appendix Table~\ref{tab:methods-common-training}.

We use Muon\cite{jordan2024muon} for hidden weight matrices and AdamW\cite{loshchilov2019adamw} for
the elemental input projection, embeddings, energy head and non-matrix
parameters, similar to the hybrid optimizer setup in DPA4\cite{li2026dpa4}. In Muon, Nesterov momentum, Newton--Schulz orthogonalization
and matrix-size scaling\cite{liu2025muon} are combined, and updates are damped using an
exponential moving average of momentum--gradient alignment\cite{joo2026surprisingeffectivenessmaskingupdates,li2026dpa4} (Appendix Note~1).
Biases, normalization gains, embeddings and elemental readout scales and shifts are excluded from weight decay.
In each training stage, we use a learning-rate schedule with linear warmup, a constant-rate phase and cosine decay, together with gradient clipping and an exponential moving average of model parameters.

\subsection{Benchmark definitions and comparisons}
\label{sec:methods-benchmarks}

\noindent{\bfseries Matbench Discovery.}
We use the benchmark's default combined performance score (CPS), molecular-dynamics score (CMDS) and diatomic score (CDS)\cite{matbench2026snapshot}.
CPS combines stability-classification F1 on unique WBM prototypes\cite{wang2021wbm},
StructureMatcher RMSD\cite{ong2013pymatgen} and thermal-conductivity symmetric
relative mean error\cite{riebesell2025matbench,pota2024thermal}.
CMDS combines discrepancies in vibrational spectra, angular distributions
and pressure distributions from DynaMat trajectories with runtime
\cite{gawkowski2026dynamat}. CDS combines energy/force accuracy, diatomic
geometry, curve shape and runtime.
Appendix Note~3 reports the component subscores and the full public comparison cohort.

For Fig.~\ref{fig:facto-overview}a and Table~\ref{tab:facto-discovery}, we
summarize overall standing by the sum of CPS, CMDS and CDS ranks.
For each metric, we rank models with complete CPS, CMDS and CDS scores in the public default cohort accessed on 22 October 2026 together with the four FACTO-OAM variants.
We order the sums of the three ranks from smallest to largest.
The resulting comparison contains 42 public models and four FACTO-OAM variants (46 complete models).
Table~\ref{tab:facto-discovery} reports the same 46-model ranks
alongside unranked FACTO-MPA scores.

\noindent{\bfseries MatCalc.}
We evaluate equilibrium formation energy and structural-fingerprint distance
(972 materials\cite{wang2021wbm}), Voigt--Reuss--Hill bulk and shear moduli (3,953 materials\cite{dejong2015elastic}),
harmonic heat capacity at 300~K (1,170 materials\cite{loew2025phonons,togo2023phonopy}), and force-softening slope
(979 materials)\cite{kaplan2025matpes,matcalc2026software,matcalc2026data,deng2025softening}.
Formation energies are evaluated relative to elemental references, and the
dimensionless structural-fingerprint distance is computed from CrystalNN
environment descriptors\cite{zimmermann2020crystalnn}.
We compute mean absolute errors (MAEs) in the predicted properties across all successful evaluations.
Full property values and baseline comparisons are reported in Appendix Note~3.

\noindent{\bfseries MLIP Arena.}
We compare diatomic, equation-of-state, energy--volume, heating/compression
and hydrogen-combustion diagnostics\cite{chiang2025arena,arena2026snapshot}.
We evaluate conservation deviation by comparing predicted forces with
numerical energy derivatives; tortuosity measures excess curve variation, and the force-flip count has an ideal value of one.
The area under the stability curve (AUC) measures finite-energy survival (with a discrete 100-bin scorer ceiling of $65/66\approx0.9848$ for complete survival), and the combustion residual serves as an operational potential-energy diagnostic.
Appendix Note~3 reports the complete set of diagnostics across these tasks.

\subsection{Inference measurements and numerical fidelity}
\label{sec:methods-inference}

\noindent{\bfseries Acceleration ablation.}
We compare plain FP32 AOTI, custom aggregation, BF16$\times$3 and their combination for the FACTO-XL model.
For each system size, we perform five warmup calls followed by 20 timed evaluations.

\noindent{\bfseries Multi-model speed comparison.}
Fourteen implementations (the four FACTO backbones and ten public baselines) are timed at the Atomic Simulation Environment (ASE)\cite{larsen2017ase} calculator-call boundary using silicon supercells with $8n^3$ atoms, $n=1,\ldots,20$.
We report mean latencies and standard deviations from direct ASE calculator-call measurements.
For FACTO, we report plain FP32 AOTI throughputs at 1,000 atoms (Fig.~\ref{fig:facto-overview}b and Table~\ref{tab:facto-discovery}) and throughputs obtained with custom aggregation and BF16$\times$3 at 8,000 atoms (Table~\ref{tab:facto-discovery}).
The scaling curves in Fig.~\ref{fig:facto-overview}c switch from plain FP32 AOTI to custom aggregation with BF16$\times$3 at each backbone's measured crossover size ($N\geq512$, $1,728$, $2,744$ and $4,096$ atoms for FACTO-XL, FACTO-L, FACTO-M and FACTO-S, respectively).
Runtime dispatch overhead is excluded.
OAM and MPA models of the same size have identical architectures and inference latencies.
For each public model, we use the fastest available implementation in the speed comparison.
Model-specific precision, acceleration backends, compilation settings
and software versions for the public comparators are listed in Appendix
Table~\ref{tab:si-speed-execution-settings}.

\noindent{\bfseries Silicon NVE trajectories.}
To isolate the effect of matrix-multiplication arithmetic on trajectory conservation, we evaluate strict FP32, BF16$\times$3 and TF32 using the same FACTO-OAM-M checkpoint with custom aggregation disabled.
For each arithmetic mode, we perform three independent runs of a 64-atom diamond-silicon supercell with a lattice constant of 5.43~\AA.
In each run, we equilibrate the system for 2~ps at a target temperature of 300~K using a Langevin thermostat with a friction coefficient of 0.01~fs$^{-1}$ under the corresponding arithmetic mode, then collect 500~ps of fixed-cell velocity-Verlet NVE dynamics at a 1~fs timestep.
Energy is sampled every ten steps, including both endpoints.

For atom count $N$ and $N_{\rm s}=50,001$ samples of total energy per atom
$u_k=(E_{{\rm potential},k}+E_{{\rm kinetic},k})/N$ at time $t_k$,
we fit $u_k=a+bt_k$ by ordinary least squares and report the signed drift $b$
and the detrended root-mean-square (RMS) fluctuation:
\begin{equation}
\mathrm{RMS}_{\rm detrended}
=\left[\frac{1}{N_{\rm s}}\sum_{k=1}^{N_{\rm s}}(u_k-a-bt_k)^2\right]^{1/2}.
\label{eq:methods-nve-rms}
\end{equation}
Both quantities are evaluated over all $N_{\rm s}$ samples, and we summarize the three detrended RMS values for each mode by their geometric mean.
We plot total-energy changes without subtracting the fitted drift.

\section*{Author contributions}
S.L. led the project and contributed to all aspects of the work, including
designing and implementing the model and acceleration methods, conducting the
experiments and writing the manuscript. C.H. implemented acceleration methods
and the training and evaluation pipeline. S.G. designed an earlier version of
the model. Y.Y. implemented acceleration methods. T.Z., Y.X., K.Y. and S.X.
contributed to discussions. W.Y. contributed to model design and project
administration.

\bibliographystyle{unsrtnat}
\begingroup
\raggedright
\bibliography{main}
\endgroup

\clearpage
\beginappendix
\begingroup
\AppendixPresentation
\setcounter{equation}{0}
\setcounter{figure}{0}
\setcounter{table}{0}
\renewcommand{\theequation}{A\arabic{equation}}
\renewcommand{\thefigure}{A\arabic{figure}}
\renewcommand{\thetable}{A\arabic{table}}
\renewcommand{\theHequation}{A\arabic{equation}}
\renewcommand{\theHfigure}{A\arabic{figure}}
\renewcommand{\theHtable}{A\arabic{table}}

\section*{Appendix Note 1: Model and training specifications}
\label{sec:appendix-note-1}

Appendix Tables~\ref{tab:methods-model-settings}--\ref{tab:methods-common-training}
specify the architecture and training hyperparameters.

\begin{table}[!htbp]
\centering
\AppendixTableStyle
\caption{{\bfseries Benchmark-model hyperparameters.}
$C_\ell$ is the number of channels for angular order $\ell$, starting at
$\ell=1$. Tensor Transport denotes the optional transport increment; when it is
disabled, the tensor stream, Geometric Tensor Injection and Tensor-to-scalar
Feedback are retained. Only scalar features are updated in the final block.}
\label{tab:methods-model-settings}
\begin{tabular*}{\linewidth}{@{\extracolsep{\fill}}lcccc@{}}
\toprule
Setting & XL & L & M & S \\
\midrule
Scalar width $D$ & 2,048 & 1,024 & 768 & 384 \\
Attention groups $G$ & 4 & 4 & 6 & 3 \\
Tensor channels $(C_1,\ldots,C_L)$ & $(512,256,128,64,32,16)$ & $(256,128,64,32)$ & $(128,64,32,16)$ & $(128,64,32,16)$ \\
Maximum angular order $L$ & 6 & 4 & 4 & 4 \\
Radial basis size $N_r$ & 64 & 64 & 64 & 32 \\
Tensor Transport enabled & Yes & Yes & No & No \\
\midrule
Interaction blocks $K$ & \multicolumn{4}{c}{5} \\
Learned cutoff $R_{\rm c}$ & \multicolumn{4}{c}{6~\AA} \\
Short-range radius $R_{\rm s}$ & \multicolumn{4}{c}{1~\AA} \\
Repulsive cutoff $R_{\rm rep}$ & \multicolumn{4}{c}{6~\AA} \\
Maximum neighbors per atom & \multicolumn{4}{c}{65,536} \\
Gaussian width $s_r$ & \multicolumn{4}{c}{$2.5/(N_r-1)$ (fixed)} \\
Residual increment scale $\alpha$ & \multicolumn{4}{c}{$40^{-1/2}$ (fixed)} \\
\bottomrule
\end{tabular*}
\end{table}

\begin{table}[!htbp]
\centering
\AppendixTableStyle
\caption{{\bfseries Stage-specific training hyperparameters.}
The three stages are MPA training from scratch,
OMat24 pretraining and OAM finetuning.
The same stage recipes apply to all four FACTO backbones (S, M, L and XL).
The atom limits are global.}
\label{tab:methods-training-settings}
\begin{tabular*}{\linewidth}{@{\extracolsep{\fill}}lccc@{}}
\toprule
Setting & MPA training & OMat24 pretraining & OAM finetuning \\
\midrule
Data & MPtrj + sAlex & OMat24 & MPtrj + sAlex \\
Initialization & Random & Random & Pretrained checkpoint \\
Maximum epochs & 100 & 20 & 40 \\
Maximum atoms & 32,000 & 64,000 & 16,000 \\
Muon base LR & $2\times10^{-3}$ & $2\times10^{-3}$ & $3\times10^{-4}$ \\
AdamW base LR & $5\times10^{-4}$ & $5\times10^{-4}$ & $7\times10^{-5}$ \\
Weight decay & $10^{-2}$ & $10^{-2}$ & $3\times10^{-4}$ \\
Final decay fraction $f_d$ & 0.35 & 0.50 & 0.35 \\
TF32 matmul in training & No & Yes & No \\
\bottomrule
\end{tabular*}
\end{table}

\begin{table}[!htbp]
\centering
\AppendixTableStyle
\caption{{\bfseries Training settings shared across the three stages.}
Weight decay in Appendix Table~\ref{tab:methods-training-settings} applies to the
eligible Muon and AdamW groups. The learned elemental energy scale and shift
are excluded from weight decay. EMA denotes exponential moving
average.}
\label{tab:methods-common-training}
\begin{tabular*}{\linewidth}{@{\extracolsep{\fill}}ll@{}}
\toprule
Setting & Value \\
\midrule
Loss weights $(w_E,w_F,w_\sigma)$ & $(1,1,1)$ \\
Muon momentum & 0.99; Nesterov enabled; no momentum warmup \\
Newton--Schulz iterations & 5 \\
Update multiplier $\gamma_{\rm opt}$ & 0.18 \\
Alignment damping & Enabled; $s_{\min}=0.1$, $\tau=2$, $\beta_a=0.9$ \\
AdamW $(\beta_1,\beta_2)$ / $\epsilon$ & $(0.9,0.95)$ / $10^{-10}$ \\
Warmup fraction & 0.01 of total steps \\
Initial / final LR multiplier & 0.2 / 0.01 \\
Gradient clipping value & 10 \\
Parameter EMA decay & 0.999 \\
\texttt{torch.compile} in training & Yes \\
\bottomrule
\end{tabular*}
\end{table}
\FloatBarrier

\noindent{\bfseries Elemental descriptors and references.}
The fixed elemental descriptor has 31 components encoding scaled elemental
properties and chemical-group indicators. MPtrj isolated-atom DFT energy references
are used in MPA training and OAM finetuning, whereas OMat24 references are used in pretraining.
We obtain the MPtrj values from Kaplan's isolated-atom energy dataset\cite{kaplan2024isolatedenergies}
and the OMat24 values from the FAIRChem repository\cite{fairchem2025repository}.

\noindent{\bfseries Muon alignment damping.}
We apply continuous damping to Muon\cite{jordan2024muon} updates using smoothed
momentum--gradient cosine alignment, similar to the deterministic alignment damping in DPA4\cite{li2026dpa4} and the alignment-based update modulation in Magma\cite{joo2026surprisingeffectivenessmaskingupdates}.
For a matrix parameter, let $g_t$ be its gradient and let
$M_t=\mu M_{t-1}+g_t$, $M_0=0$, be its momentum buffer. The Nesterov
input to Newton--Schulz is $g_t+\mu M_t$. We calculate
\begin{equation}
  \begin{aligned}
    \zeta_t&=\beta_a\zeta_{t-1}+(1-\beta_a)
      \max\!\left(0,\frac{\langle M_t,g_t\rangle_F}
        {\max(\|M_t\|_F,\varepsilon_a)
         \max(\|g_t\|_F,\varepsilon_a)}\right),\\
    \delta_t&=s_{\min}+(1-s_{\min})\zeta_t^\tau,
  \end{aligned}
  \label{eq:supp-muon-damping}
\end{equation}
with $\zeta_0=1$ and $\varepsilon_a=10^{-12}$.
We multiply the orthogonalized update by $\delta_t\gamma_{\rm opt}\sqrt{\max(n_{\rm row},n_{\rm col})}$ before applying the learning rate (similar to the matrix-size scaling in refs.~\cite{liu2025muon,li2026dpa4}), where $n_{\rm row}$ and $n_{\rm col}$ are the numbers of rows and columns of the parameter matrix.
Alignment is evaluated in FP32.
Decoupled weight decay is applied separately and is not multiplied by $\delta_t$.

\noindent{\bfseries Learning-rate schedule.}
We use a warmup--stable--decay (WSD) schedule\cite{hu2024minicpmunveilingpotentialsmall}.
For $T$ optimizer steps, the warmup lasts
$T_w=\lceil0.01T\rceil$ steps and the final cosine decay lasts
$T_d=\lceil f_dT\rceil$, where $f_d$ is the stage-specific fraction in
Appendix Table~\ref{tab:methods-training-settings}.
Between these phases, the learning-rate multiplier is one.
During warmup, the multiplier at optimizer step $t$ is $0.2+0.8t/T_w$; during decay, it is
\begin{equation}
  0.01+\frac{0.99}{2}
    \left[1+\cos\!\left(\pi\frac{t-(T-T_d)}{T_d}\right)\right].
\end{equation}
The scheduler is updated at each optimizer step over the configured training horizon.

\noindent{\bfseries Dataset partitions and validation reuse.}
In MPA training and OAM finetuning, we concatenate the configuration-level
MPtrj fitting partition with sAlex without dataset reweighting.
In OMat24 pretraining, we use the official training, validation and test splits.
Atom-count batching is used across all stages.

\section*{Appendix Note 2: Acceleration and numerical fidelity}
\label{sec:appendix-note-2}

\noindent{\bfseries Three-product bfloat16 FP32 emulation (BF16$\times$3).}
Let $\mathcal B$ round an FP32 matrix to BF16. For operands $\mathbf X$ and
$\mathbf W$, we split off the leading residual terms\cite{henry2019bfloat16}:
\begin{equation}
  \begin{aligned}
  \mathbf X_{\rm hi}&=\mathcal B(\mathbf X),&
  \mathbf X_{\rm lo}&=\mathcal B(\mathbf X-\mathbf X_{\rm hi}),\\
  \mathbf W_{\rm hi}&=\mathcal B(\mathbf W),&
  \mathbf W_{\rm lo}&=\mathcal B(\mathbf W-\mathbf W_{\rm hi}).
  \end{aligned}
  \label{eq:facto-bf16-split}
\end{equation}
Residual subtraction and accumulation are performed in FP32, and three tensor-core products
are used to approximate the original multiplication:
\begin{equation}
  \mathbf X\mathbf W\simeq
  \mathbf X_{\rm hi}\mathbf W_{\rm hi}
  +\mathbf X_{\rm hi}\mathbf W_{\rm lo}
  +\mathbf X_{\rm lo}\mathbf W_{\rm hi}.
  \label{eq:facto-bf16x3}
\end{equation}
The residual--residual product is omitted.
We apply this FP32 emulation to matrix multiplications during energy evaluation and differentiation.

\begin{table}[!htbp]
\centering
\AppendixTableStyle
\caption{{\bfseries Energy conservation for all silicon trajectories.}
Each 64-atom diamond-Si trajectory runs for 500~ps with a 1~fs timestep.
$T_0$ is the instantaneous temperature after 2~ps of equilibration at a 300~K target.
Drift is the fitted total-energy slope; RMS is calculated after subtracting that line.
Both quantities are evaluated over all samples using FACTO-OAM-M.}
\label{tab:nve-all-seeds}
\begin{tabular*}{\linewidth}{@{\extracolsep{\fill}}lcrrr@{}}
\toprule
Arithmetic & Run & {$T_0$ (K)} & {Drift ($\mu$eV\,atom$^{-1}$\,ps$^{-1}$)} & {RMS ($\mu$eV\,atom$^{-1}$)} \\
\midrule
FP32 & 1 & \num{265.8} & \num{-0.007223} & \num{4.250} \\
FP32 & 2 & \num{247.8} & \num{-0.02029} & \num{3.957} \\
FP32 & 3 & \num{303.5} & \num{-0.007017} & \num{4.648} \\
\midrule
BF16$\times$3 & 1 & \num{265.8} & \num{0.04072} & \num{4.079} \\
BF16$\times$3 & 2 & \num{247.8} & \num{-0.01555} & \num{3.950} \\
BF16$\times$3 & 3 & \num{303.5} & \num{-0.02458} & \num{4.406} \\
\midrule
TF32 & 1 & \num{266.1} & \num{0.1739} & \num{28.98} \\
TF32 & 2 & \num{247.8} & \num{-0.1712} & \num{31.18} \\
TF32 & 3 & \num{303.3} & \num{0.1729} & \num{35.40} \\
\bottomrule
\end{tabular*}
\end{table}

\begin{figure}[!htbp]
  \centering
  \includegraphics[width=0.72\linewidth]{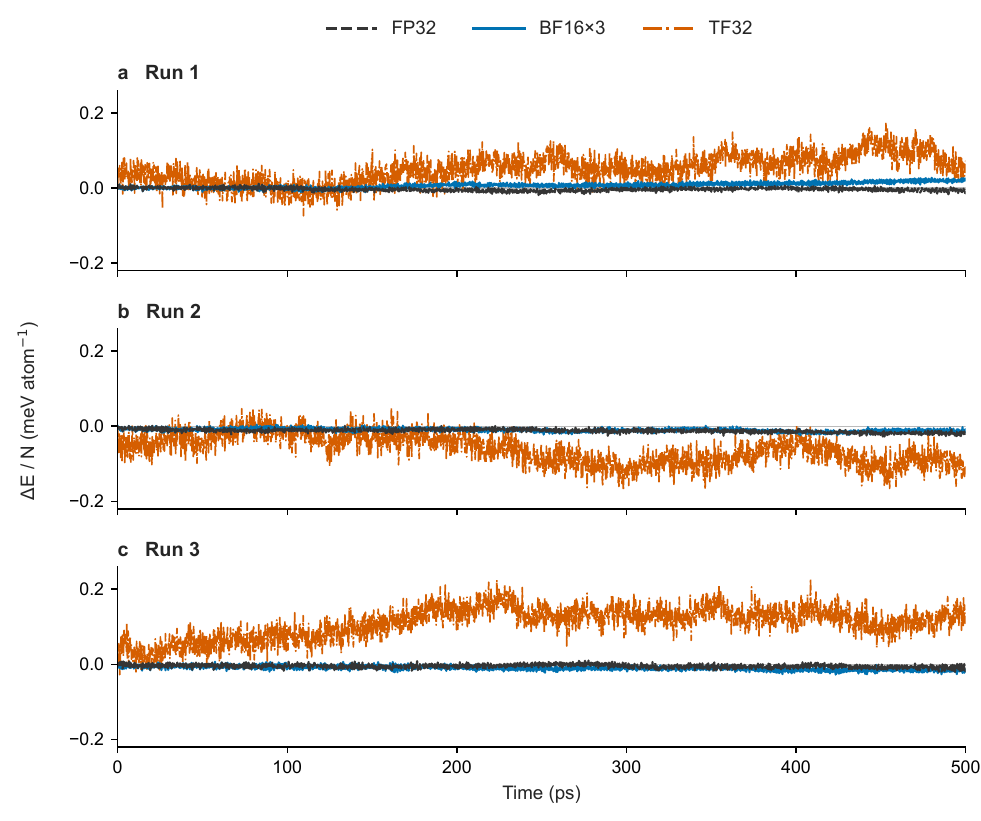}
  \caption{{\bfseries Total-energy changes in all silicon NVE runs.}
  Panels a--c show three independently prepared runs for FP32 (dashed),
  BF16$\times$3 (solid) and TF32 (dash-dotted). The vertical axis is
  $[E_{\rm total}(t)-E_{\rm total}(0)]/N$, where total energy is potential
  plus kinetic energy and $N=64$. Each trajectory spans 500~ps with a 1~fs
  timestep using FACTO-OAM-M. No fitted linear drift is subtracted.
  Traces are unsmoothed at 0.2~ps spacing;
  drift and detrended RMS in Appendix Table~\ref{tab:nve-all-seeds} use all
  samples at 0.01~ps spacing.}
  \label{fig:nve-traces}
\end{figure}

\FloatBarrier
\Needspace{24\baselineskip}\section*{Appendix Note 3: Benchmark measurements and comparison sources}\label{sec:appendix-note-3}

\subsection*{Matbench Discovery}
\par\addvspace{12pt}\noindent\begin{minipage}{\linewidth}\AppendixTableStyle
\captionsetup{type=table,hypcap=false}\caption{{\bfseries CPS and its subscores for FACTO models.} All quantities are dimensionless. F1 is evaluated on unique WBM prototypes; RMSD is evaluated at a symmetry tolerance of $10^{-2}$.}\label{tab:si-mbd-core}
\begin{tabular*}{\linewidth}{@{\extracolsep{\fill}}lrrrrr@{}}
\toprule
Model & CPS & F1 & RMSD & $\kappa_{\rm SRME}$ & $\kappa_{\rm SRE}$ \\
\midrule
\multicolumn{6}{@{}l@{}}{{\bfseries FACTO-OAM}}\\
XL & \num{0.890} & \num{0.9219} & \num{0.05980} & \num{0.1560} & \num{0.06490} \\
L & \num{0.883} & \num{0.9178} & \num{0.06158} & \num{0.1733} & \num{0.07037} \\
M & \num{0.863} & \num{0.9036} & \num{0.06533} & \num{0.2284} & \num{0.1225} \\
S & \num{0.855} & \num{0.8964} & \num{0.06634} & \num{0.2425} & \num{0.1167} \\
\addlinespace[4pt]
\multicolumn{6}{@{}l@{}}{{\bfseries FACTO-MPA}}\\
XL & \num{0.867} & \num{0.9101} & \num{0.06550} & \num{0.2218} & \num{0.09135} \\
L & \num{0.859} & \num{0.8988} & \num{0.06355} & \num{0.2415} & \num{0.1003} \\
M & \num{0.847} & \num{0.8915} & \num{0.06574} & \num{0.2753} & \num{0.1271} \\
S & \num{0.832} & \num{0.8834} & \num{0.06543} & \num{0.3296} & \num{0.1902} \\
\bottomrule\end{tabular*}
\end{minipage}\par\addvspace{8pt}

\par\addvspace{12pt}\noindent\begin{minipage}{\linewidth}\AppendixTableStyle
\captionsetup{type=table,hypcap=false}\caption{{\bfseries CMDS and its subscores for FACTO models.} Errors are normalized percentage discrepancies. $S_t$ is the runtime subscore.}\label{tab:si-mbd-md}
\begin{tabular*}{\linewidth}{@{\extracolsep{\fill}}lrrrrrr@{}}
\toprule
Model & CMDS & vDOS (\%) & ADF (\%) & Pressure (\%) & Time ($10^3$ s) & $S_t$ \\
\midrule
\multicolumn{7}{@{}l@{}}{{\bfseries FACTO-OAM}}\\
XL & \num{0.751} & \num{22.60} & \num{1.313} & \num{46.29} & \num{17.91} & \num{0.8038} \\
L & \num{0.773} & \num{22.78} & \num{1.447} & \num{48.41} & \num{10.81} & \num{0.9478} \\
M & \num{0.789} & \num{23.01} & \num{1.368} & \num{46.56} & \num{7.771} & \num{1.000} \\
S & \num{0.792} & \num{23.46} & \num{1.377} & \num{44.97} & \num{7.533} & \num{1.000} \\
\addlinespace[4pt]
\multicolumn{7}{@{}l@{}}{{\bfseries FACTO-MPA}}\\
XL & \num{0.756} & \num{23.29} & \num{1.407} & \num{44.51} & \num{17.41} & \num{0.8118} \\
L & \num{0.775} & \num{23.24} & \num{1.424} & \num{47.49} & \num{10.78} & \num{0.9484} \\
M & \num{0.793} & \num{22.99} & \num{1.458} & \num{45.18} & \num{7.778} & \num{1.000} \\
S & \num{0.790} & \num{23.65} & \num{1.523} & \num{45.32} & \num{7.433} & \num{1.000} \\
\bottomrule\end{tabular*}
\end{minipage}\par\addvspace{8pt}

\par\addvspace{12pt}\noindent\begin{minipage}{\linewidth}\AppendixTableStyle
\captionsetup{type=table,hypcap=false}\caption{{\bfseries Diatomic PBE-reference errors.} Energy and well depth are in eV, force in eV/\AA, and distances in \AA.}\label{tab:si-mbd-diatomic-errors}
\begin{tabular*}{\linewidth}{@{\extracolsep{\fill}}lrrrrr@{}}
\toprule
Model & Energy & Force & Wall distance & Bond length & Well depth \\
\midrule
\multicolumn{6}{@{}l@{}}{{\bfseries FACTO-OAM}}\\
XL & \num{0.6657} & \num{0.6338} & \num{0.07888} & \num{0.2384} & \num{0.6193} \\
L & \num{1.065} & \num{0.7017} & \num{0.08888} & \num{0.2868} & \num{1.223} \\
M & \num{0.7499} & \num{0.6323} & \num{0.06709} & \num{0.1831} & \num{0.7709} \\
S & \num{0.7915} & \num{0.6896} & \num{0.06755} & \num{0.1469} & \num{0.7332} \\
\addlinespace[4pt]
\multicolumn{6}{@{}l@{}}{{\bfseries FACTO-MPA}}\\
XL & \num{0.8904} & \num{0.9282} & \num{0.09832} & \num{0.1910} & \num{0.6883} \\
L & \num{0.8260} & \num{0.8752} & \num{0.09117} & \num{0.1484} & \num{0.6382} \\
M & \num{0.8439} & \num{0.8168} & \num{0.1155} & \num{0.1328} & \num{0.6937} \\
S & \num{0.8966} & \num{1.177} & \num{0.08129} & \num{0.1518} & \num{0.6342} \\
\bottomrule\end{tabular*}
\end{minipage}\par\addvspace{8pt}

\par\addvspace{12pt}\noindent\begin{minipage}{\linewidth}\AppendixTableStyle
\captionsetup{type=table,hypcap=false}\caption{{\bfseries Diatomic physicality and runtime.} Flips and tortuosity are dimensionless.}\label{tab:si-mbd-diatomic-physicality}
\begin{tabular*}{\linewidth}{@{\extracolsep{\fill}}lrrrrr@{}}
\toprule
Model & Jump (eV) & Flips & Tortuosity & Time (s) & CDS \\
\midrule
\multicolumn{6}{@{}l@{}}{{\bfseries FACTO-OAM}}\\
XL & \num{0.4266} & \num{3.529} & \num{1.014} & \num{91.43} & \num{0.771} \\
L & \num{0.1988} & \num{2.460} & \num{1.027} & \num{66.95} & \num{0.781} \\
M & \num{0.1652} & \num{2.678} & \num{1.010} & \num{53.03} & \num{0.822} \\
S & \num{0.2249} & \num{2.540} & \num{1.011} & \num{52.21} & \num{0.825} \\
\addlinespace[4pt]
\multicolumn{6}{@{}l@{}}{{\bfseries FACTO-MPA}}\\
XL & \num{0.3900} & \num{2.839} & \num{1.023} & \num{87.84} & \num{0.770} \\
L & \num{0.2118} & \num{2.644} & \num{1.015} & \num{66.66} & \num{0.801} \\
M & \num{0.1508} & \num{2.425} & \num{1.017} & \num{52.28} & \num{0.813} \\
S & \num{0.1573} & \num{2.310} & \num{1.009} & \num{51.95} & \num{0.807} \\
\bottomrule\end{tabular*}
\end{minipage}\par\addvspace{8pt}

\begingroup\AppendixTableStyle
\begin{longtable}{@{}>{\raggedright\arraybackslash}p{0.70\linewidth}rrr@{}}
\caption{{\bfseries Public Matbench Discovery comparison at 19 October 2026.} The 42 scored default-cohort models are ordered by CPS\cite{matbench2026snapshot}.}\label{tab:si-mbd-public}\\
\toprule
Public model & CPS & CMDS & CDS \\
\midrule\endfirsthead
\caption[]{{\bfseries Public Matbench Discovery comparison at 19 October 2026 (continued).}}\\
\toprule
Public model & CPS & CMDS & CDS \\
\midrule\endhead
\midrule\multicolumn{4}{r}{\textit{Continued on next page}}\\\endfoot
\bottomrule\endlastfoot
Prophet-OAME-MBD & \num{0.912} & \num{0.578} & \num{0.572} \\
TECE-OAM-RRA-1.0\cite{xu2026tece} & \num{0.908} & \num{0.630} & \num{0.270} \\
EquFlashV2 & \num{0.907} & \num{0.651} & \num{0.644} \\
EquiformerV3+DeNS-OAM & \num{0.902} & \num{0.638} & \num{0.269} \\
GRACE-3L-OAM-L & \num{0.900} & \num{0.739} & \num{0.699} \\
PET-OAM-XL & \num{0.898} & \num{0.634} & \num{0.388} \\
TACE-OAM-L & \num{0.889} & \num{0.728} & \num{0.761} \\
eSEN-30M-OAM & \num{0.888} & \num{0.625} & \num{0.648} \\
EquFlash & \num{0.888} & \num{0.678} & \num{0.328} \\
NequIP-OAM-XL & \num{0.886} & \num{0.623} & \num{0.796} \\
MatRIS-10M-OAM\cite{zhou2026matris} & \num{0.877} & \num{0.581} & \num{0.678} \\
SevenNet-Omni-i12 & \num{0.873} & \num{0.626} & \num{0.696} \\
NequIP-OAM-L & \num{0.870} & \num{0.670} & \num{0.818} \\
GRACE-2L-OAM-L & \num{0.865} & \num{0.759} & \num{0.694} \\
ORB v3 & \num{0.860} & \num{0.736} & \num{0.834} \\
DPA-4.0.1-Pro-MPtrj & \num{0.840} & \num{0.598} & \num{0.123} \\
Allegro-OAM-L & \num{0.840} & \num{0.602} & \num{0.828} \\
GRACE-2L-OAM & \num{0.837} & \num{0.751} & \num{0.527} \\
EquiformerV3+DeNS-MP & \num{0.830} & \num{0.622} & \num{0.413} \\
DPA-3.1-3M-FT & \num{0.802} & \num{0.647} & \num{0.533} \\
eSEN-30M-MP & \num{0.797} & \num{0.593} & \num{0.417} \\
MACE-MPA-0 & \num{0.795} & \num{0.659} & \num{0.728} \\
MatRIS-10M-MP & \num{0.778} & \num{0.593} & \num{0.264} \\
AlphaNet-v1-OAM & \num{0.769} & \num{0.653} & \num{0.739} \\
MatterSim v1 5M & \num{0.767} & \num{0.659} & \num{0.717} \\
GRACE-1L-OAM & \num{0.761} & \num{0.749} & \num{0.552} \\
Eqnorm MPtrj & \num{0.756} & \num{0.639} & \num{0.711} \\
Nequix MP PFT & \num{0.755} & \num{0.616} & \num{0.565} \\
NequIP-MP-L & \num{0.730} & \num{0.629} & \num{0.573} \\
Nequix MP & \num{0.729} & \num{0.603} & \num{0.561} \\
Allegro-MP-L & \num{0.720} & \num{0.556} & \num{0.747} \\
SevenNet-l3i5 & \num{0.714} & \num{0.651} & \num{0.533} \\
HIENet & \num{0.707} & \num{0.653} & \num{0.649} \\
GRACE-2L-MPtrj & \num{0.681} & \num{0.698} & \num{0.355} \\
BAM-MP-DeNS & \num{0.672} & \num{0.564} & \num{0.708} \\
MACE-MP-0 & \num{0.637} & \num{0.633} & \num{0.627} \\
eqV2 M & \num{0.558} & \num{0.595} & \num{0.283} \\
ORB v2 MPA & \num{0.528} & \num{0.750} & \num{0.678} \\
eqV2 S DeNS & \num{0.522} & \num{0.615} & \num{0.258} \\
ORB v2 MPtrj & \num{0.470} & \num{0.727} & \num{0.667} \\
M3GNet & \num{0.428} & \num{0.593} & \num{0.747} \\
CHGNet & \num{0.400} & \num{0.554} & \num{0.297} \\
\end{longtable}\endgroup

\begin{figure}[!htbp]
\centering
\includegraphics[width=\linewidth]{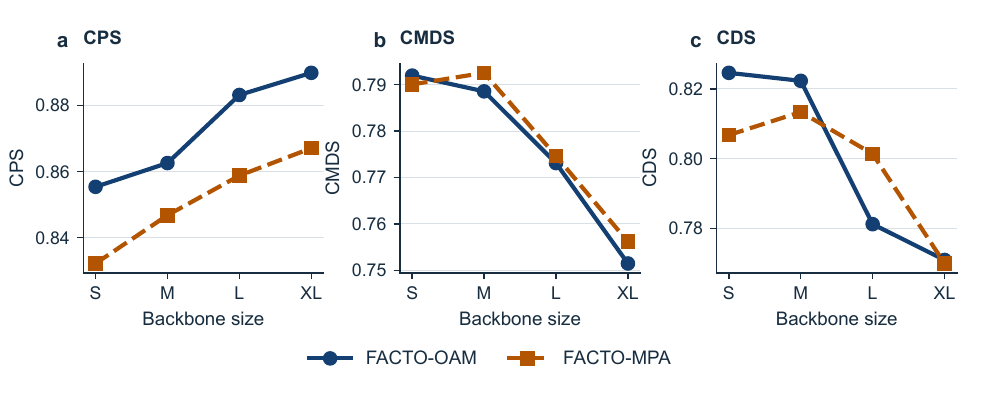}
\caption{{\bfseries Effect of training data and model size on Matbench Discovery scores.}
Panels a--c show CPS, CMDS and CDS for OAM (solid circles) and MPA (dashed squares).
All scores are higher-is-better. Training data are defined in
Appendix Note~1; component measurements are reported in the accompanying tables.}
\label{fig:si-training-settings}
\end{figure}
\FloatBarrier
\Needspace{24\baselineskip}\subsection*{MatCalc}
\par\addvspace{12pt}\noindent\begin{minipage}{\linewidth}\AppendixTableStyle
\captionsetup{type=table,hypcap=false}\caption{{\bfseries Crystalline-property errors and force calibration.} Formation energy is in meV/atom, elastic moduli in GPa and $C_V$ in J\,mol$^{-1}$\,K$^{-1}$. Structural distance $d$ and force slope are dimensionless; the ideal slope is one. Published baselines are from ref.~\citenum{kavanagh2026foundation}.}\label{tab:si-matcalc-values}
\begin{tabular*}{\linewidth}{@{\extracolsep{\fill}}lrrrrrr@{}}
\toprule
Model & $d$ & $E_f$ & $K_{\rm VRH}$ & $G_{\rm VRH}$ & $C_V$ & Force slope \\
\midrule
\multicolumn{7}{@{}l@{}}{{\bfseries FACTO-OAM}}\\
XL & \num{0.1959} & \num{22.35} & \num{11.05} & \num{12.39} & \num{4.397} & \num{0.9861} \\
L & \num{0.2117} & \num{25.38} & \num{11.71} & \num{12.04} & \num{4.801} & \num{0.9847} \\
M & \num{0.2102} & \num{24.73} & \num{12.56} & \num{13.49} & \num{4.978} & \num{0.9798} \\
S & \num{0.2291} & \num{26.06} & \num{12.58} & \num{13.54} & \num{5.654} & \num{0.9758} \\
\addlinespace[4pt]\multicolumn{7}{@{}l@{}}{{\bfseries Published baselines}}\\
NequIP-OAM-XL & \num{0.2790} & \num{25.40} & \num{12.70} & \num{14.10} & \num{4.020} & \num{0.9790} \\
NequIP-OAM-L & \num{0.2810} & \num{26.60} & \num{14.90} & \num{14.90} & \num{4.490} & \num{0.9730} \\
Allegro-OAM-L & \num{0.2870} & \num{24.70} & \num{12.70} & \num{14.70} & \num{4.380} & \num{0.9690} \\
MatterSim-5M & \num{0.3220} & \num{40.50} & \num{16.90} & \num{15.80} & \num{5.460} & \num{0.9730} \\
UMA-s & \num{0.2670} & \num{25.00} & \num{14.90} & \num{21.50} & \num{3.600} & \num{0.9860} \\
\bottomrule\end{tabular*}
\end{minipage}\par\addvspace{8pt}

\FloatBarrier\Needspace{20\baselineskip}\subsection*{MLIP Arena}
\par\addvspace{12pt}\noindent\begin{minipage}{\linewidth}\AppendixTableStyle
\captionsetup{type=table,hypcap=false}\caption{{\bfseries Homonuclear-diatomic behavior on MLIP Arena.} Spearman rank correlations $\rho$, flips and tortuosity are dimensionless.}\label{tab:si-arena-diatomics}
\begin{tabular*}{\linewidth}{@{\extracolsep{\fill}}lrrrr@{}}
\toprule
Diagnostic & OAM-XL & OAM-L & OAM-M & OAM-S \\
\midrule
Conservation (eV/\AA) & \num{0.01707} & \num{0.01360} & \num{0.01685} & \num{0.01426} \\
$\rho(E,\,\mathrm{repulsion})$ & \num{-0.9823} & \num{-0.9915} & \num{-0.9970} & \num{-0.9985} \\
$\rho(F,\,\mathrm{descending})$ & \num{-0.7081} & \num{-0.9666} & \num{-0.9053} & \num{-0.9052} \\
Tortuosity & \num{1.020} & \num{1.023} & \num{1.010} & \num{1.011} \\
Energy jump (eV) & \num{0.01374} & \num{0.006905} & \num{0.006975} & \num{0.007406} \\
Force flips & \num{4.219} & \num{2.708} & \num{3.062} & \num{2.833} \\
$\rho(E,\,\mathrm{attraction})$ & \num{0.8562} & \num{0.5403} & \num{0.6985} & \num{0.7496} \\
$\rho(F,\,\mathrm{ascending})$ & \num{0.6491} & \num{0.5643} & \num{0.6273} & \num{0.7387} \\
\bottomrule\end{tabular*}
\end{minipage}\par\addvspace{8pt}

\par\addvspace{12pt}\noindent\begin{minipage}{\linewidth}\AppendixTableStyle
\captionsetup{type=table,hypcap=false}\caption{{\bfseries Heating and compression stability on MLIP Arena.} AUC integrates finite-energy survival (discrete 100-bin ceiling $65/66\approx0.9848$); exponents give the power-law scaling of steps per second with atom count.}\label{tab:si-arena-stability}
\begin{tabular*}{\linewidth}{@{\extracolsep{\fill}}lrrrr@{}}
\toprule
Diagnostic & OAM-XL & OAM-L & OAM-M & OAM-S \\
\midrule
Heating AUC & \num{0.9848} & \num{0.9848} & \num{0.9799} & \num{0.9848} \\
Compression AUC & \num{0.9848} & \num{0.9848} & \num{0.9848} & \num{0.9848} \\
Heating exponent & \num{0.5579} & \num{0.5340} & \num{0.4225} & \num{0.2854} \\
Compression exponent & \num{0.6479} & \num{0.6442} & \num{0.5227} & \num{0.3860} \\
\bottomrule\end{tabular*}
\end{minipage}\par\addvspace{8pt}

\par\addvspace{12pt}\noindent\begin{minipage}{\linewidth}\AppendixTableStyle
\captionsetup{type=table,hypcap=false}\caption{{\bfseries Reactive hydrogen combustion.} The enthalpy residual is Arena's potential-energy diagnostic; COM denotes center of mass.}\label{tab:si-arena-combustion}
\begin{tabular*}{\linewidth}{@{\extracolsep{\fill}}lrrrr@{}}
\toprule
Diagnostic & OAM-XL & OAM-L & OAM-M & OAM-S \\
\midrule
Enthalpy residual (kcal/mol) & \num{3.814} & \num{1.878} & \num{3.619} & \num{5.288} \\
COM drift (\AA) & \num{0.001330} & \num{0.001764} & \num{0.001375} & \num{0.001247} \\
Throughput (steps\,s$^{-1}$) & \num{45.41} & \num{73.89} & \num{123.4} & \num{157.7} \\
Water yield (\%) & \num{86.98} & \num{88.80} & \num{90.10} & \num{85.94} \\
\bottomrule\end{tabular*}
\end{minipage}\par\addvspace{8pt}

\par\addvspace{12pt}\noindent\begin{minipage}{\linewidth}\AppendixTableStyle
\captionsetup{type=table,hypcap=false}\caption{{\bfseries Relaxed equation-of-state diagnostics.} Cells give the mean; $\rho$ denotes Spearman's rank correlation. All quantities are dimensionless.}\label{tab:si-arena-eos_bulk}
\begin{tabular*}{\linewidth}{@{\extracolsep{\fill}}lrrrr@{}}
\toprule
Diagnostic & OAM-XL & OAM-L & OAM-M & OAM-S \\
\midrule
Derivative flips & \num{1.039} & \num{1.048} & \num{1.047} & \num{1.041} \\
Tortuosity & \num{1.006} & \num{1.009} & \num{1.007} & \num{1.006} \\
$\rho(E,\,\mathrm{compression})$ & \num{-0.9978} & \num{-0.9986} & \num{-0.9981} & \num{-0.9995} \\
$\rho(\Delta E,\,\mathrm{compression})$ & \num{0.9953} & \num{0.9965} & \num{0.9964} & \num{0.9968} \\
$\rho(E,\,\mathrm{tension})$ & \num{0.9847} & \num{0.9843} & \num{0.9811} & \num{0.9828} \\
\bottomrule\end{tabular*}
\end{minipage}\par\addvspace{8pt}

\par\addvspace{12pt}\noindent\begin{minipage}{\linewidth}\AppendixTableStyle
\captionsetup{type=table,hypcap=false}\caption{{\bfseries Fixed-coordinate energy--volume diagnostics.} Cells give the mean; $\rho$ denotes Spearman's rank correlation. All quantities are dimensionless.}\label{tab:si-arena-ev}
\begin{tabular*}{\linewidth}{@{\extracolsep{\fill}}lrrrr@{}}
\toprule
Diagnostic & OAM-XL & OAM-L & OAM-M & OAM-S \\
\midrule
Derivative flips & \num{1.006} & \num{1.008} & \num{1.008} & \num{1.011} \\
Tortuosity & \num{1.011} & \num{1.000} & \num{1.001} & \num{1.001} \\
$\rho(E,\,\mathrm{compression})$ & \num{-0.9983} & \num{-0.9983} & \num{-0.9983} & \num{-0.9983} \\
$\rho(\Delta E,\,\mathrm{compression})$ & \num{0.9995} & \num{0.9999} & \num{1.000} & \num{0.9999} \\
$\rho(E,\,\mathrm{tension})$ & \num{0.9960} & \num{0.9991} & \num{0.9987} & \num{0.9989} \\
\bottomrule\end{tabular*}
\end{minipage}\par\addvspace{8pt}
\FloatBarrier

\Needspace{45\baselineskip}
\subsection*{Third-party inference settings}
\par\addvspace{6pt}\noindent\begin{minipage}{\linewidth}\AppendixTableStyle
\setlength{\tabcolsep}{4pt}
\captionsetup{type=table,hypcap=false}
\caption{{\bfseries Third-party inference and acceleration settings.}
Settings are for the ten public comparators in the main speed comparison.
All weights are stored in FP32. OEQ and cuEq denote OpenEquivariance and
cuEquivariance, respectively; AMP and UQ denote automatic mixed precision
and uncertainty estimation.}
\label{tab:si-speed-execution-settings}
\begin{tabularx}{\linewidth}{@{}>{\raggedright\arraybackslash}p{0.24\linewidth}>{\raggedright\arraybackslash}X>{\raggedright\arraybackslash}p{0.25\linewidth}@{}}
\toprule
Model & Execution and acceleration & Software and revision \\
\midrule
Prophet-OAME-MBD & OEQ tensor-product kernels enabled; graph compilation disabled.
& Prophet \texttt{dcfdc0b}; PyTorch 2.8.0; OEQ 0.7.0 \\
\addlinespace[4pt]
EquiformerV3+DeNS-OAM & Conservative gradient checkpoint; eager inference; graph compilation, AMP and TF32 disabled.
& EquiformerV3 \texttt{a7300c58}; PyTorch 2.7.1+cu128; e3nn 0.5.6 \\
\addlinespace[4pt]
GRACE-3L-OAM-L & TensorFlow/XLA; dense \texttt{uniform} aggregation; UQ disabled.
& TensorPotential 0.6.1; TensorFlow 2.16.2; tf-keras 2.16.0 \\
\addlinespace[4pt]
NequIP-OAM-XL & AOTInductor + OEQ; compiled \texttt{.nequip.pt2} model loaded for inference.
& NequIP 0.19.1; PyTorch 2.8.0; OEQ 0.7.0 \\
\addlinespace[4pt]
NequIP-OAM-L & AOTInductor + OEQ; compiled \texttt{.nequip.pt2} model loaded for inference.
& NequIP 0.19.1; PyTorch 2.8.0; OEQ 0.7.0 \\
\addlinespace[4pt]
TACE-OAM-L & OEQ enabled; \texttt{float32} evaluation.
& TACE 0.2.1; PyTorch 2.8.0; OEQ 0.5.4 \\
\addlinespace[4pt]
eSEN-30M-OAM & Conservative gradient inference in eager mode; AMP and TF32 disabled.
& fairchem-core 1.10.0; PyTorch 2.4.1; e3nn 0.6.0 \\
\addlinespace[4pt]
ORB v3 & Conservative; internal graph network compiled with \texttt{torch.compile} (default mode, dynamic shapes); fixed brute-force neighbor search.
& orb-models 0.5.5; PyTorch 2.8.0 \\
\addlinespace[4pt]
MACE-MPA-0 & Medium checkpoint; cuEq enabled; \texttt{float32} evaluation.
& MACE 0.3.16; PyTorch 2.8.0; cuEq 0.11.0 \\
\addlinespace[4pt]
GRACE-2L-OAM-L & FP32 export supporting dense \texttt{uniform} aggregation; TensorFlow/XLA; UQ disabled.
& TensorPotential 0.6.1; TensorFlow 2.16.2; tf-keras 2.16.0 \\
\bottomrule
\end{tabularx}
\end{minipage}\par\addvspace{8pt}

\subsection*{Matbench Discovery score definitions}

We use the default Matbench Discovery scorer and the public model cohort
accessed on 22 October 2026\cite{matbench2026snapshot}.
With $[x]_0^1=\min(1,\max(0,x))$, the combined performance score is
\begin{equation}
\mathrm{CPS}=0.5\,\mathrm{F1}
+0.1[1-r/0.15]_0^1+0.4(1-\kappa_{\rm SRME}/2).
\label{eq:methods-cps}
\end{equation}
Here, F1 measures stability classification on unique WBM prototypes,
$r$ is the dimensionless StructureMatcher RMSD at a symmetry tolerance of
$10^{-2}$, and $\kappa_{\rm SRME}$ is the thermal-conductivity symmetric
relative mean error\cite{riebesell2025matbench,pota2024thermal}.
The SRME resolves mode-level errors, whereas $\kappa_{\rm SRE}$ compares
total thermal conductivities; both are dimensionless.
The public comparison includes active models that predict forces: 43
registered models, of which 42 have complete CPS, CMDS and CDS values.

For molecular dynamics, let $e_v$, $e_a$ and $e_p$ be normalized percentage
discrepancies in vibrational density of states, angular distributions and
pressure distributions across the 17 DynaMat trajectories (each 80,000 steps of 0.25~fs)\cite{gawkowski2026dynamat}.
The score is
\begin{equation}
\begin{aligned}
\mathrm{CMDS}&=0.3(1-e_v/100)+0.2(1-e_a/100)\\
&\quad+0.3(1-e_p/100)+0.2S_t,\\
S_t&=\left[1-\frac{\log(t/9000)}{\log(300000/9000)}\right]_0^1,
\end{aligned}
\label{eq:methods-cmds}
\end{equation}
where $t$ is total rollout time in seconds and $S_t$ is the runtime subscore reported in Appendix Table~\ref{tab:si-mbd-md}.
For $e_v$, the Wasserstein-1 distance between normalized spectra is divided
by the reference spectral standard deviation after clipping the reference
tail at 0.999 cumulative power.
For $e_a$, the Wasserstein-1 distance is
divided by the reference distribution's distance to a structureless
$\sin\theta$ background. These two ratios are capped at one and converted
to percentages.
The pressure error $e_p$ measures histogram non-overlap: 0\% for identical distributions and 100\% for disjoint distributions.

The diatomic score CDS combines four pillars with weights $4/9$ for
energy/force accuracy, $2/9$ for geometry, $2/9$ for physical curve shape
and $1/9$ for runtime.
For a quantity $x$ with ideal floor $a$ and baseline $b$, the linear subscore is $[(b-x)/(b-a)]_0^1$; for runtime, we instead use $[\log(b/x)/\log(b/a)]_0^1$.
Appendix Table~\ref{tab:cds-definition} gives the component weights and normalization endpoints.
The weighted sum of the pillar scores is multiplied by $1-n_{\rm excluded}/87$; no elements are excluded for FACTO, and 79 PBE reference curves are used for reference-relative errors.

\begin{table}[!htbp]
\centering\AppendixTableStyle
\caption{{\bfseries CDS component weights and normalization endpoints.} Weights are
normalized within each pillar.}
\label{tab:cds-definition}
\begin{tabular*}{\linewidth}{@{\extracolsep{\fill}}llrrl@{}}
\toprule
Pillar & Quantity & Weight & Floor & Baseline \\
\midrule
Accuracy & Energy MAE & $5/9$ & 0 & 4 eV \\
 & Force MAE & $4/9$ & 0 & 5 eV/\AA \\
Geometry & Wall-radius MAE & $1/2$ & 0 & 0.5 \AA \\
 & Bond-length error & $1/3$ & 0 & 0.8 \AA \\
 & Well-depth error & $1/6$ & 0 & 3.5 eV \\
Physicality & Energy jump & $2/5$ & 0 & 3 eV \\
 & Force reversals & $2/5$ & 1 & 3.5 \\
 & Tortuosity & $1/5$ & 1 & 2 \\
Speed & Runtime (logarithmic) & 1 & 50 s & 5,000 s \\
\bottomrule
\end{tabular*}
\end{table}

\FloatBarrier
\subsection*{MatCalc observable conventions}

The equilibrium, elasticity, phonon and force-softening collections contain
972, 3,953, 1,170 and 979 materials (9,308 high-energy configurations), respectively, corresponding to the WBM equilibrium, binary Materials Project elasticity, binary Alexandria phonon and WBM high-energy collections\cite{matcalc2026data}.
Equilibrium observables are formation-energy MAE and dimensionless
CrystalNN structural-fingerprint distance; elastic moduli use the Voigt--Reuss--Hill average.
We report harmonic heat capacity at 300~K using the benchmark's Phonopy cell-molar convention\cite{kaplan2025matpes,matcalc2026software,deng2025softening}, and force calibration as the mean proportional-fit slope between predicted and DFT force components across materials (ideal value one).
Baseline values in Appendix Table~\ref{tab:si-matcalc-values} are taken from Table~3 of ref.~\citenum{kavanagh2026foundation}.

\subsection*{MLIP Arena diagnostic definitions}

Public MLIP Arena values are from the 27 March 2026 comparison
\cite{chiang2025arena,arena2026snapshot}.
On homonuclear diatomics and volumetric deformation scans (Appendix Tables~\ref{tab:si-arena-diatomics}, \ref{tab:si-arena-eos_bulk} and~\ref{tab:si-arena-ev}), conservation deviation compares predicted forces with numerical energy derivatives on the sampled grid, tortuosity measures excess curve variation relative to a single-well curve, and $\rho$ denotes Spearman's rank correlation coefficient.
In Appendix Table~\ref{tab:si-arena-stability}, heating and compression AUCs integrate finite-energy trajectory survival using the benchmark's 100-edge discrete bin scorer (which has a maximum achievable value of $(97+1/2)/99=65/66\approx0.9848$ when all trajectories survive to the final step), and the heating and compression scaling exponents $\alpha_{\rm speed}$ are obtained by fitting steps per second to a power law proportional to $N^{-\alpha_{\rm speed}}$, where $N$ is the atom count.
In Appendix Table~\ref{tab:si-arena-combustion}, the combustion enthalpy residual is computed from the potential-energy change divided by the stoichiometric yield of 128 water molecules and converted using 23.0~kcal\,mol$^{-1}$ per eV relative to $-68.3078$~kcal\,mol$^{-1}$.

\FloatBarrier
\endgroup

\end{document}